\documentstyle[twocolumn,epsfig,aps]{revtex}
\draft	

\begin{document}
\twocolumn[\hsize\textwidth\columnwidth\hsize\csname @twocolumnfalse\endcsname
\title{Ordering Dynamics of Heisenberg Spins
with Torque\,: Crossover, Spin waves and Defects}
\author{Jayajit Das$^{1,2}$\cite{JAY} and Madan Rao$^2$\cite{MAD}}

\address{$^1$Institute of Mathematical Sciences, Taramani, Chennai
600113,
India\\
$^2$Raman Research Institute, C.V. Raman Avenue,
Sadashivanagar, Bangalore 560080,
India}

\date{\today}

\maketitle

\begin{abstract}

We study the effect of a torque induced by the local molecular field on
the phase ordering dynamics of the Heisenberg model when the total
magnetization is conserved. The torque drives the zero-temperature
ordering dynamics to a new fixed point, characterized by exponents $z=2$
and $\lambda \approx 5$. This `torque-driven' fixed point is approached
at times such that $g^2 t\gg 1$, where $g$ is the strength of the torque.
All physical quantities, like the domain size $L(t)$ and the equal and
unequal time correlation functions obey a crossover scaling form over the
entire range of $g$. An attempt to understand this crossover behavior
from the approximate Gaussian Closure Scheme fails completely, implying
that the dynamics at late times cannot be understood from the dynamics of
defects alone. We provide convincing arguments that the spin
configurations can be decomposed in terms of defects and spin-waves which
interact with each other even at late times. In the absence of the torque
term, the spin waves decay faster, but even so we find that the Gaussian
closure scheme is inconsistent. In the latter case the inconsistency may
be remedied by including corrections to a simple gaussian distribution.
For completeness we include a discussion of the ordering dynamics at
$T_c$, where the torque is shown to be relevant, with exponents
$z=4-\varepsilon/2$ and $\lambda = d$ (where $\varepsilon = 6-d$).  We
show to all orders in perturbation theory that $\lambda=d$ as a
consequence of the conservation law.

\end{abstract}

]
\vskip1.0in

\section{Introduction}

When a many-body system like a magnet or a binary fluid is quenched from
its disordered high temperature phase to its ordered configuration at low
temperatures, the slow annealing of ``defects'' (interfaces in binary
fluids, vortices (hedgehogs) in XY (Heisenberg) magnets) separating
competing domains, makes the dynamics very slow. The system organizes
itself into a self similar spatial distribution of domains characterized
by a single diverging length scale which typically grows algebraically in
time $L(t) \sim t^{1/z}$. This spatial distribution of domains is
reflected in the scaling behavior of the equal-time correlation function
$C(r,t) \sim f(r/L(t))$.  The autocorrelation function, $A(t)
\sim L(t)^{-\lambda}$ is a measure of the memory of the initial
configurations. The exponents $z$ and $\lambda$ and the scaling function
$f(x)$ characterize the dynamical universality classes at the zero
temperature fixed point (ZFP)\cite{BRAY}.

The above phenomenology suggests that the asymptotic dynamics of the order
parameter is dominated by the dynamics of its defects, and that bulk
fluctuations (concentration waves in a binary fluid, spin waves in a
magnet) relax fast and decouple from the dynamics of defects at late
times. This picture is at the heart of recent approximate theories such as
the Gaussian Closure Scheme \cite{MAZENKO1,BRAY}. 

But is this picture accurate ? In this paper we shall study the very
realistic example of the conservative dynamics of a Heisenberg magnet
driven by a torque induced by the local molecular field, and show that the
longer-lived spin waves couple to the defects even at late times, driving
the system to a new fixed point. The new `torque-driven' fixed point
characterized by $z=2$ and $\lambda \approx 5.05$, is accessed after a
crossover time $t_c \sim 1/g^2$ (where $g$ is the strength of the torque).
Crossover scaling forms describe physical quantities at late times (like
the domain size $L(t,g)$ and correlation functions $C(r,t,g)$ and
$A(t,g)$) for all values of $g$. In the absence of the torque, the
spin-waves decay faster, but even so we find that the Gaussian Closure
Scheme is internally inconsistent. This inconsistency may however be
rectified by including leading corrections to the gaussian distribution
(as suggested by Mazenko \cite{MAZENKO2} for the dynamics of the conserved
scalar (Ising) order parameter).

For completeness we also study the effects of including the torque in the
dynamics following a quench to the critical point $T_c$. As reported in
earlier studies \cite{MA}, the torque is relevant with exponents
$z=4-\varepsilon/2$ and $\lambda = d$ (where $\varepsilon = 6-d$).
We show to all orders in perturbation theory that $\lambda =d$ which
follows as a consequence of the conservation of total magnetization
\cite{SM,JANS,KISS}.

\section{Heisenberg Magnet and Precessional Dynamics}

The order parameter ${\vec \phi}$ (whose components are $\phi_{\alpha}$
with $\alpha = 1,2,3$) describing a coarse-grained spin density in a
Heisenberg ferromagnet in three dimensions experiences a torque from the
joint action of the external field (if present) and the local molecular
field. In response the spins precess with a Larmour frequency $\Omega_L$
about the total magnetic field. Coupling to various faster degrees of
freedom like lattice vibrations or electrons, causes a dissipation in
energy and an eventual relaxation towards equilibrium.

This dynamics follows from the generalized Langevin equation 
and the Poisson algebra \cite{DAS},
\begin{equation}
\frac{\partial \phi_{\alpha}}{\partial t} = \Gamma \nabla^{2}\frac
{\delta F}{\delta \phi_{\alpha}}\, + \,\Omega_L \,\epsilon_{\alpha \beta \gamma}
\,\phi_{\beta}\,\frac{\delta F}{\delta \phi_{\gamma}} + \,\eta_{\alpha}\,\,.
\label{eq:dye}
\end{equation}
The noise $\vec \eta$ arising from the heat bath has mean
zero and is conservative,
\begin{equation}
\langle \eta_{\alpha}({\bf x},t) \eta_{\beta}({\bf x}',t')\rangle = 2\,k_B
T\, \Gamma\,\delta_{\alpha \beta}\,\nabla^2 \delta({\bf x}-{\bf
x}') \delta(t-t')\,.
\label{eq:noise}
\end{equation}
The free-energy functional $F$ is taken to be of the Landau-Ginzburg form,
\begin{eqnarray}
F[\vec{\phi}] & = & \int \,d^{3}x \,\left[ \,\frac{\sigma}{2} \,(\nabla\vec{\phi})^2 
-\,\frac{r}{2}\,(\vec{\phi} \cdot \vec{\phi})\,+\,\frac{u}{4}\,(\vec{\phi} 
\cdot \vec{\phi})^2 \, \right] \, .
\label{eq:landau} 
\end{eqnarray}
The second term in Eq.\ (\ref{eq:dye}) is clearly the torque ${\vec M}
\times {\vec H}$, where ${\vec H} \equiv - \delta F/\delta {\vec \phi}$ is
the local molecular field.

Both the inertial term (by virtue of $F$ being rotationally invariant in spin
space) and the dissipation conserve the total spin, and so the full
equations of motion (\ref{eq:dye}) also conserve the total spin.

Since the noise correlator is proportional to temperature, we may drop it
in our discussion of zero temperature quenches.  We then scale space ${\bf
x}$, time $t$ and the order parameter $\vec{\phi}$ as \[ {\bf x }
\rightarrow \sqrt{\frac{r}{\sigma}}\,{\bf x} , \: \:  t \rightarrow
\frac{\Gamma r^2 t}{\sigma}, \: \: \vec \phi \rightarrow
\sqrt{\frac{u}{r}} \vec \phi \] to obtain the equation of motion in
dimensionless form,

\begin{equation}
\frac{\partial \vec{\phi}}{\partial t} = \nabla^2\left(-\nabla^2 \vec{\phi} 
\, -\, \vec{\phi} + \, (\vec{\phi} \cdot \vec{\phi})\, 
\vec{\phi}\right) \, + \,g
\, \left(\vec{\phi} \times \nabla^2\vec{\phi}\right) \, . 
\label{eq:dyeq2}
\end{equation}

The dimensionless parameter $g = ({\Omega_{L}\sigma}/{\Gamma})(r
u)^{-1/2}$ is the ratio of the precession frequency to the relaxation
rate. Setting $\Omega_L \sim 10^{7}$\,Hz and $\Gamma \sim 10^6 -
10^{10}$\,Hz, gives $g$ in the range of $\sim 10^{-3} - 10$.

\section{Phase Ordering Dynamics at $T = 0$}

Let us now prepare the system initially in the paramagnetic phase and quench
to zero temperature. We study the time evolution of the spin 
configurations as they evolve according to Eq.\ (\ref{eq:dyeq2}). We 
calculate the equal time correlator,
\begin{equation}
 C({\bf r} ,t ) \equiv \langle {\vec \phi({\bf x} ,t)} \cdot {\vec
\phi({\bf 
x+r},t)}\rangle\,,
\label{eq:eqcorr}
\end{equation}
and the autocorrelator,
\begin{equation}
C({\bf 0}, t_1=0, t_2=t) \equiv A(t) = \langle \vec \phi({\bf r},
0) \cdot \vec 
\phi({\bf r}, t) \rangle \,,
\label{eq:autocorr}
\end{equation}
where the angular brackets are averages over the random initial
conditions and space. At late times these correlators should attain their 
scaling forms
\begin{equation}
C({\bf r}, t) \sim f(r/L(t))
\label{eq:scorr}
\end{equation}
and
\begin{equation}
A(t) \sim L(t)^{-\lambda}\,.
\label{eq:sauto}
\end{equation}
The length scale $L(t)$, which is a measure of the distance between
defects, may be evaluated either from the first zero of $C(r,t)$ or from
the scaling of the energy density, $\varepsilon = \frac{1}{V}\, \int d
{\bf r} \,\langle \,(\,\nabla \vec \phi({\bf r}, t)\,)^2 \,\rangle \sim
L(t)^{-2}$, and grows with time as $L(t) \sim t^{1/z}$. We compute the
scaling function $f(x)$, the dynamical exponent $z$ and the
autocorrelation exponent $\lambda$ by simulating the Langevin Eq.\
(\ref{eq:dyeq2}).

\subsection{Langevin Simulation}  	

The Langevin simulation is performed by discretizing Eq.\ (\ref{eq:dyeq2})
on a simple cubic lattice (with size $N$ ranging from $50^3$ to $60^3$)
and adopting an Euler scheme for the derivatives \cite{SR}. The space and
time intervals have been chosen to be $\triangle x = 2.5$ and $\triangle t
= 0.2$. With this choice of parameters, we have checked that the resulting
coupled map does not lead to any instability. We have also checked that
the results remain unchanged on slight variations of $\triangle x$ and
$\triangle t$.  Throughout our simulation we have used periodic boundary
conditions.

The correlation functions Eqs.\ (\ref{eq:eqcorr}), (\ref{eq:autocorr}) are
calculated for values of $g$ ranging from $0$ to $1$. Measured quantities
are averaged over $5-10$ initial configurations. The initial
configurations are taken from two ensembles, both in the disordered phase.
In ensemble {\bf A}, $\vec{\phi}(t=0)$ is uniformly distributed within the
volume of a unit sphere centered at the origin. $\vec{\phi}$ at different
spatial points are uncorrelated.  In ensemble {\bf B}, $\vec{\phi}(t=0)$
is uniformly distributed on the surface of a unit sphere centered at the
origin. $\vec{\phi}$ at different spatial points are again uncorrelated.
We consider these two initial conditions to check if the late time
dynamics is insensitive to the choice of initial ensemble (as long as they
do not introduce any long-range correlations).

We first report simulation results for ensemble ${\bf A}$.

Figure\,$1$ is a scaling plot of $C(r,t)$ versus $r/L(t)$ for various
values of the parameter $g$, where $L(t)$ is extracted from the first zero
of $C(r,t)$. Note that the scaling function for $g=0$ is very different
from those for $g>0$\,;  further the $g>0$ scaling functions do not seem
to depend on the value of $g$. This suggests that the dynamics crosses
over to a new `torque-driven' ZFP.  This is also revealed in the values of
the dynamical exponent $z$. In Fig.\ 2, a plot of $L(t)$ versus $t$ gives
the expected value of $z=4$ when $g=0$. For $g>0$, we see a distinct
crossover from $z=4$ when $t<t_c(g)$ to $z=2$ when $t>t_c(g)$. The
crossover time $t_c(g)$ decreases with increasing $g$. The same $z$
exponent and crossover are obtained from the scaling behaviour of the
energy density $\varepsilon$.

To make sure that our results are not affected by finite size, we compute
3 relevant time scales (shown in Table 1 below) --- (1)  $t_{c}(g)$, the
crossover time from a $t^{1/4}$ to a $t^{1/2}$ growth, (2)  $t_{s}(g)$,
the time at which asymptotic scaling begins, (3)  $t_{fs}$, the time at
which finite size effects become prominent. It is clear from the Table
that $t_c < t_s < t_{fs}$, as it should be if our data is to be free of
finite size artifacts. A general rule-of-thumb is that finite size effects
start becoming prominent when the domain size gets to be of order $1/3$
the system size, and we see from Table 1 that $L_{max}/N$ is comfortably
less than $1/3$.

\vskip0.5cm
\begin{center}
Table 1
\end{center}
\begin{center}
\begin{tabular}{|l|l|l|l|l|l|} \hline \hline
$g$ & $t_{c}(g)$ & $t_{s}(g)$  &  $t_{fs}$  & $L_{max}/N$ & $f_{min}$
 \\ \hline 
$0  $ & $\,-\,$ & $900$  &  $>7650$  & $1/10$ at $t=7650$ & $-0.14$
 \\ \hline 
$0.1$ & $3150$ & $\geq 7650$  &  $>7650$  & $1/6$ at $t=7650$ & $-0.08$
 \\ \hline 
$0.3$ & $900$ & $1350$  &  $>7650$  & $1/4$ at $t=7650$ & $-0.06$
 \\ \hline
$0.5$ & $450$ & $900$  &  $4950$  & $1/3.7$ at $t=4950$ & $-0.06$
 \\ \hline 
\end{tabular}
\end{center}
\vskip0.5cm

\begin{figure}
\centerline{\epsfig{figure=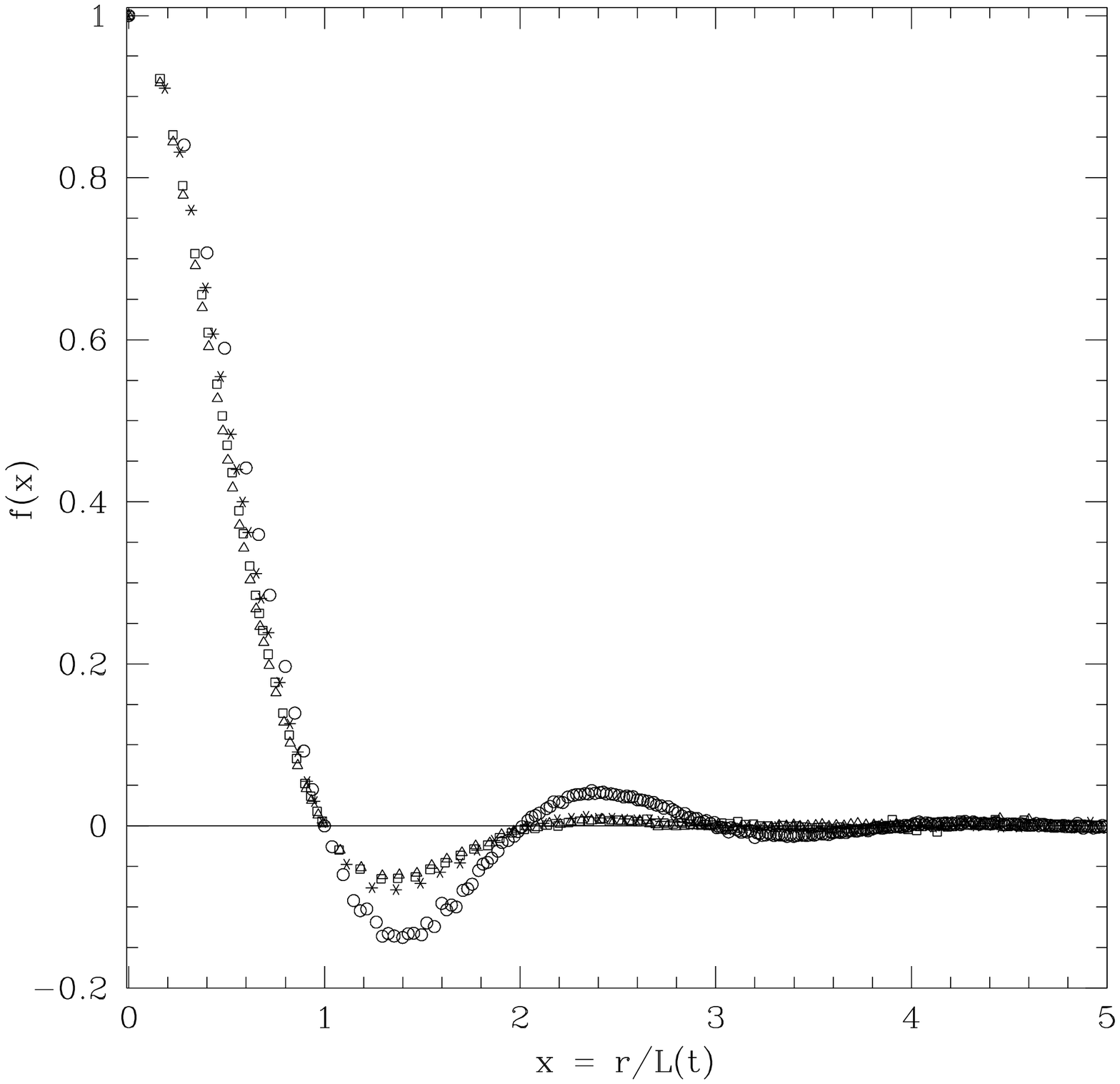,width=7.0cm,height=7.0cm}}
\end{figure} 
FIG. 1. Scaling plot of $C(r,t)$ for $N=50^3$. 
The scaling function $f(x)$ changes as $g$
is varied from $g=0\,(\circ)$ to $g\neq0\,
(g=0.1(\ast),\,0.3(\triangle),\,0.5(\Box))$. \\

\begin{figure}
\centerline{\epsfig{figure=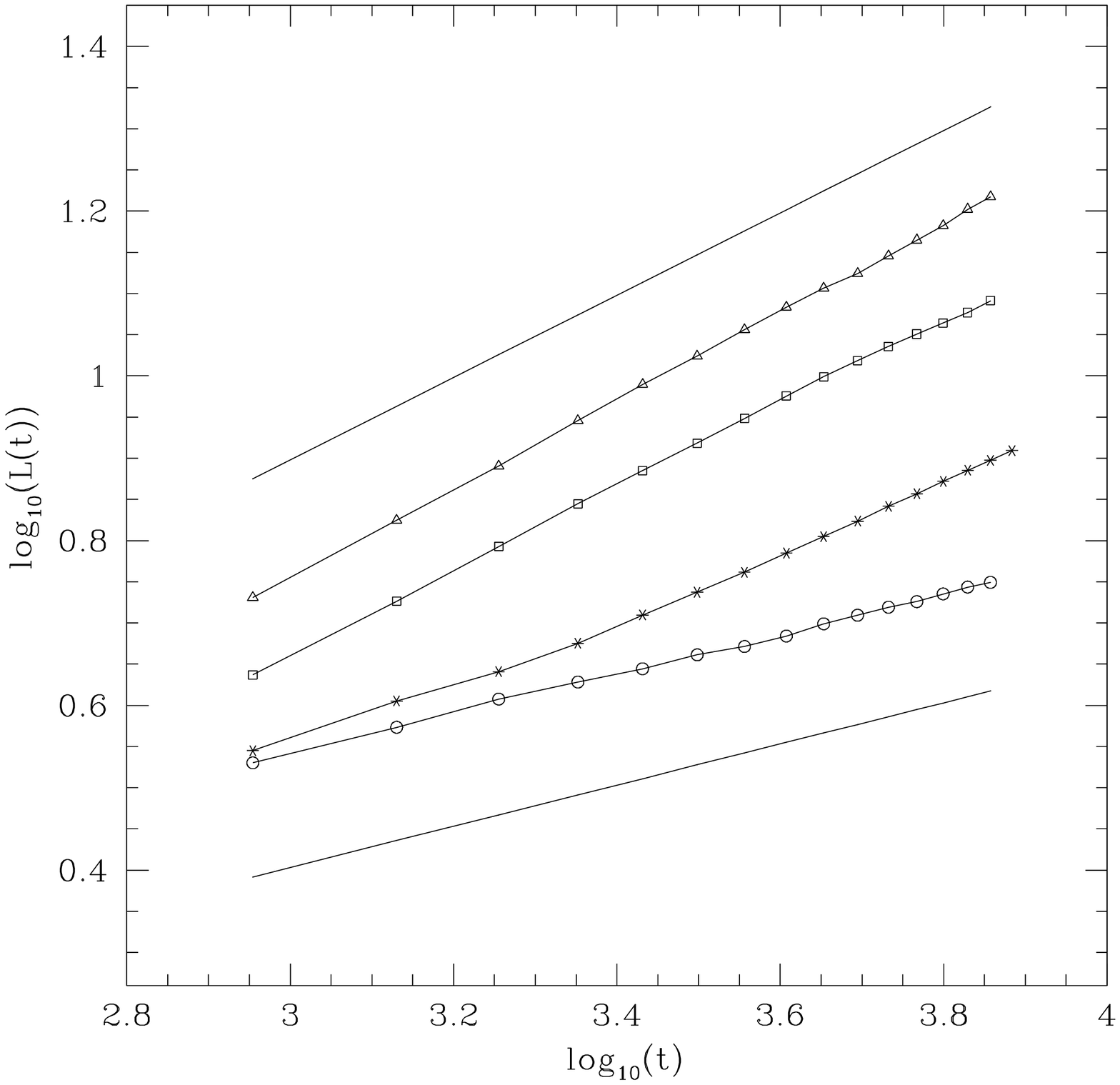,width=7.0cm,height=7.0cm}}
\end{figure} 

FIG. 2. Log-Log plot of $L(t)$. 
At $g=0\,(\circ)$ we find that  $z=4$ (line of slope $0.25$ 
drawn at the bottom
for comparison). At $g\neq0\, (g=0.1(\ast),\,0.3(\Box),\,
0.5(\triangle))$, $z$ crosses over from 
$4$ to $2$ (line of slope $0.5$
drawn at the top).
\\
\\

The last column in Table 1 shows $f_{min}$, the value of the scaling
function evaluated at the first minimum as a function of $g$. It is easy
to see why $f_{min}(g) < f_{min}(g=0)$, since the precession of the spins
about the local molecular field would cause spins from neighboring
"domains" to be less anti-correlated. This is borne out by computing the
spin-wave correction to an approximate form of $C(r,t\,; g=0)$ (given in
Eq.\ (\ref{eq:hyper}), more on this later) to quadratic order in the
spin-wave amplitude \cite{DAS}.

The autocorrelation function $A(t)$ is calculated for $g=0, 0.2$ and $0.3$
(Fig.\ 3). The simulations have been done on a lattice of size $60^3$ and
averaged over $10$ initial configurations (we have to average over a large
number of initial configurations for smoother data). The $\lambda$
exponent extracted from the asymptotic decay of $A(t)$ clearly suggests a
crossover from $\lambda =2.2$ to $\lambda\approx 5.05$. The numerical
determination of $\lambda$ is subject to large errors\cite{YRD,DAS} and is
very sensitive to finite size effects, and so we have to go to very late
times and hence large system sizes to obtain accurate results.

\begin{figure}
\centerline{\epsfig{figure=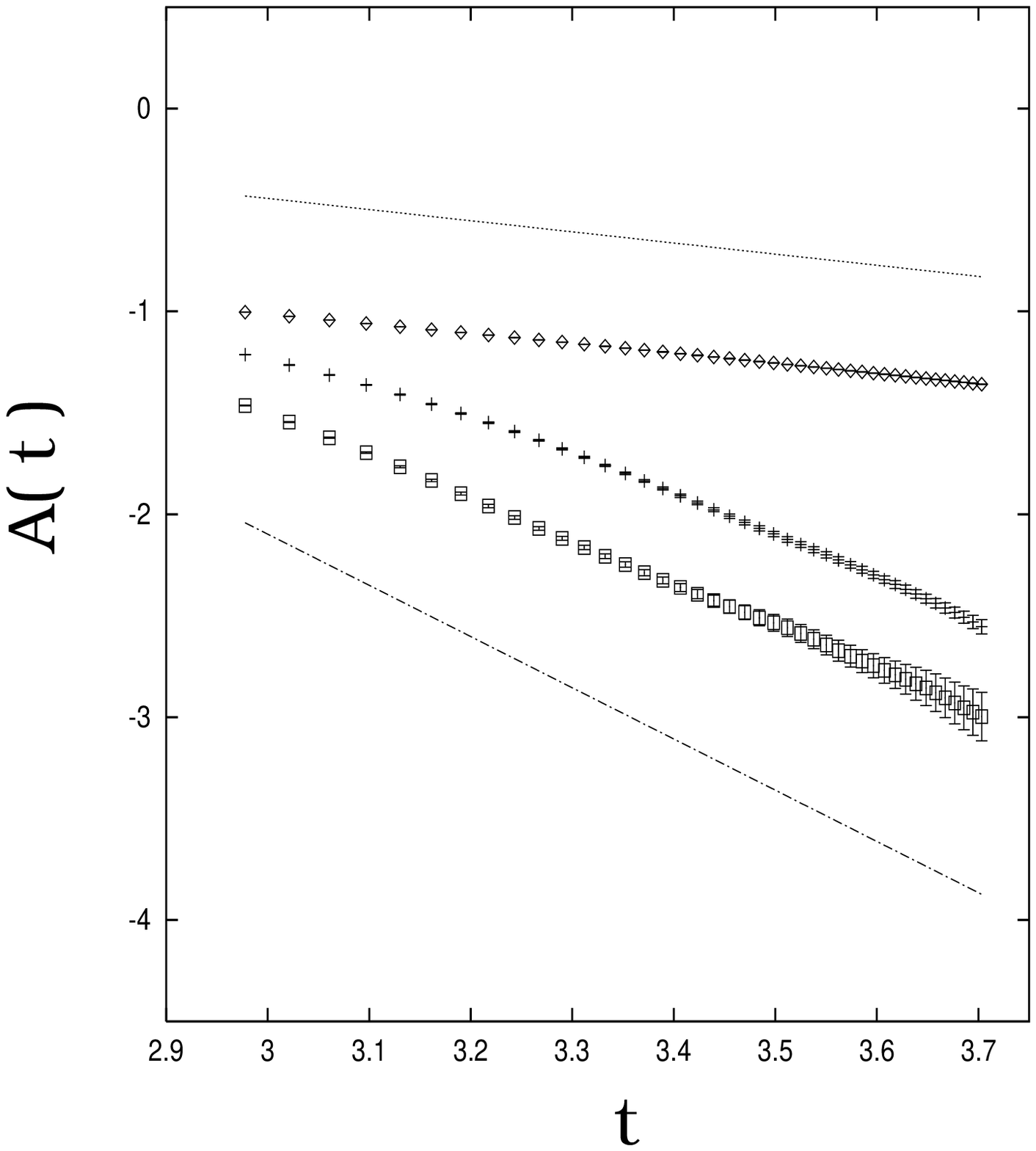,width=7.0cm,height=7.0cm}}
\end{figure} 
FIG. 3. Log-Log plot of $A(t)$ vs $t$ for $g=0(\diamond),\,0.2(+),\,
0.3(\Box)$. Solid line on top has the form $a/t^{\lambda/z}$ where
$\lambda=2.19$ and $z=4$ (corresponding to the $g=0$ fixed point) while
the one below has a $\lambda= 5.05$ and $z=2$ (corresponding to the
`torque-driven' fixed point). 
\\
\\	

To make sure that we collect asymptotic data untainted by finite size, we
compute two time scales (Table 2) --- (i) $t_{fit}(g)$,
the time beyond which $A(t)$ can be fit with a power law
$a\,(t+t_{0})^{-\lambda/z}$, (ii) $t_{fs}$, the time at which finite size
effects on $A(t)$ become prominent. The crossover time $t_c$ was displayed
in Table 1.

To determine $t_{fs}$ we plot an effective exponent $\lambda_{eff} = - t d
(\log A(t))/dt $ as a function of $1/t$. The derivative is calculated
numerically with a $\delta t = 15$ (in units of the time discretisation
$\Delta t$). We see from Fig.\ 4, that at late times $t>t_{fs}$,
$\lambda_{eff}$ crosses over to being a decreasing function of time,
clearly a finite size effect. This estimate of $t_{fs}$ is not very
sensitive to the choice of $\delta t$ changing by $1\%$ (for $g=0.2$) and
$3.5 \%$ (for $g=0.3$) as $\delta t$ changes by $5$ units. Note that
finite size effects in $A(t)$ appear earlier than in $C(r,t)$.

It is seen from Table 2 that $t_{fit} < t_{fs}$, as it should if we are to
have an accurate determination of $\lambda$.

\vskip0.5cm
\begin{center}
Table 2
\end{center}
\begin{center}
\begin{tabular}{|l|l|l|l|} \hline \hline
$g$ &$t_{fit}(g)$ &  $t_{fs}$  & $\lambda$
 \\ \hline 
$0.0$ & $900$ & $ >9000 $ &  $2.199\pm 7.5 \times 10^{-3}$
 \\ \hline 
$0.2$  & $1500$ & $5376$ & $5.100 \pm 6.1 \times 10^{-3}$
 \\ \hline 
$0.3$ & $900$ &  $5181$  &   $5.010 \pm 2.3\times 10^{-3}$
 \\ \hline
\end{tabular}
\end{center}
\vskip0.5cm

\begin{figure}
\centerline{\epsfig{figure=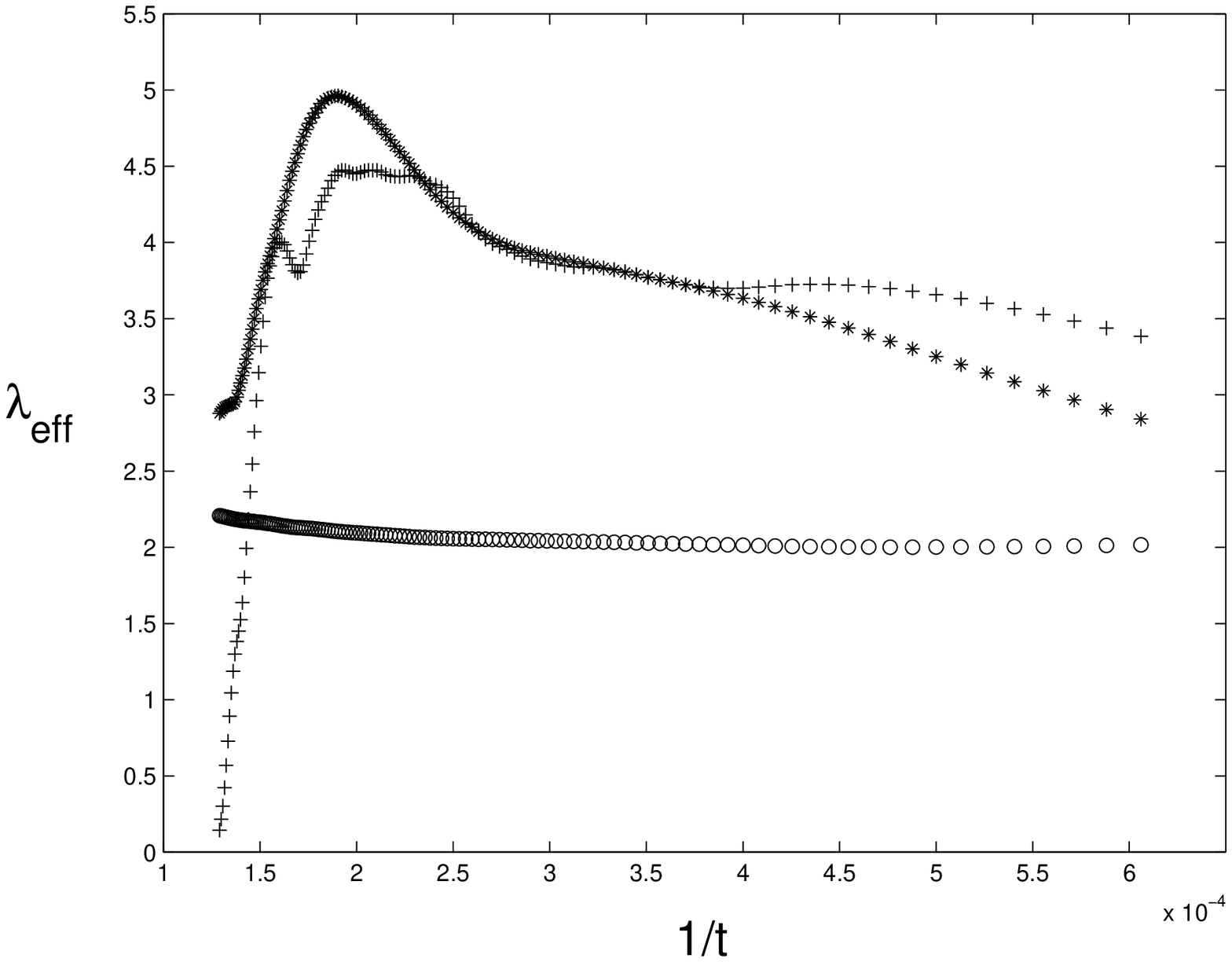,width=7.0cm,height=7.0cm}}
\end{figure} 
FIG. 4.  $\lambda_{eff}$ versus $1/t$ for
$g=0.0(\circ),\,0.2(\ast),\,0.3(+)$.  Finite size effects set in
when $\lambda_{eff}$ starts becoming a decreasing function of time. For
$g=0$ we do not see any finite size effects in $\lambda$ within our
simulation times.
\\
\\
The last column of Table 2 lists the value of $\lambda$ as a function of
$g$. The data presented and the plot in Fig.\ 3 clearly support a
crossover from $\lambda=2.2$ at $g=0$ to $\lambda=5.05$ at $g\neq0$.
The values of $\lambda$ satisfy the bound derived in \cite{YRD}.

We now present results of the Langevin simulation for initial conditions
taken from ensemble {\bf B}. We find that the value of $z$, the form
of the scaling functions $f(x)$ (Fig.\ 5) and the decay of the
autocorrelation function $A(t)$ (Fig.\ 6) are insensitive to the choice of
initial conditions.

\begin{figure}
\centerline{\epsfig{figure=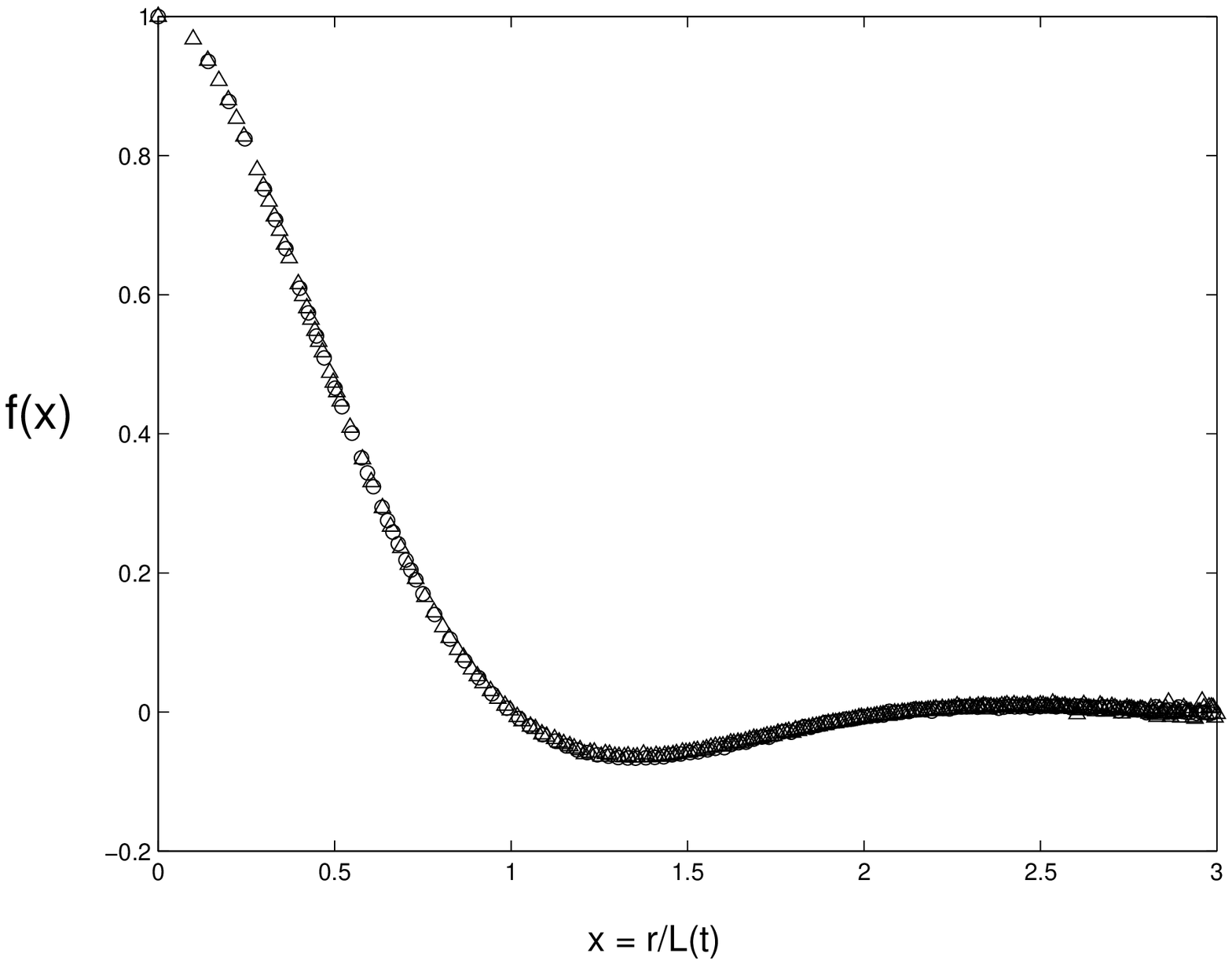,width=7.0cm,height=7.0cm}}
\end{figure} 
FIG. 5. Scaling function $f(x)$ vs $x$ for $g=0.3$ using
ensembles ${\bf A}(\circ)$ and ${\bf B}(\triangle)$.
\\
\\ 

Since the initial condition {\bf B} sets the magnitude of the
spins to its $T=0$ equilibrium value, the crossover time $t_c$ is smaller
than for ensemble {\bf A}. For the same reason the domain sizes computed
using ensemble {\bf B} are larger than that of {\bf A}.

\begin{figure}
\centerline{\epsfig{figure=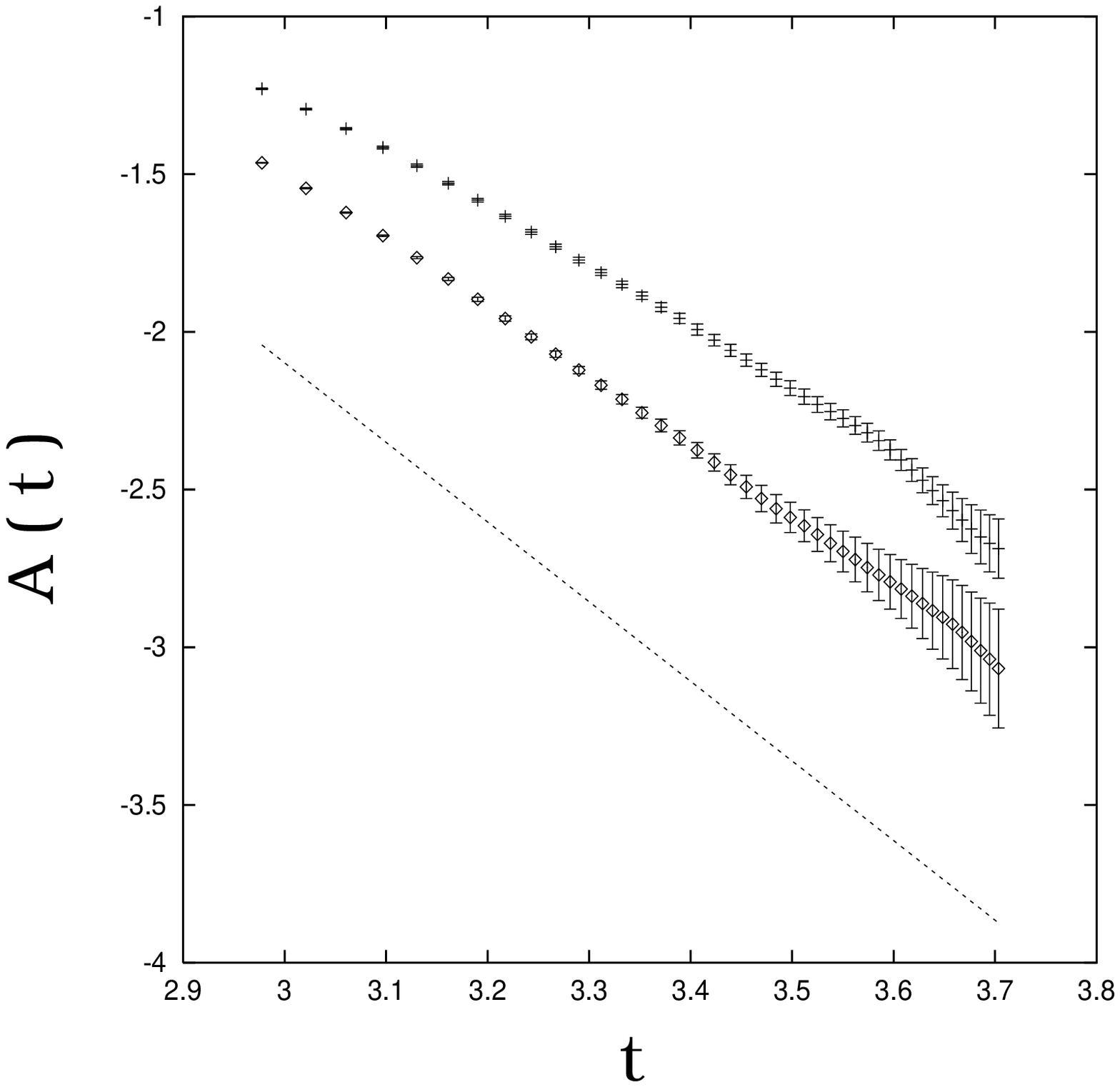,width=7.0cm,height=7.0cm}}
\end{figure} 
FIG. 6. Log-Log plot of $A(t)$ for $g=0.3$ using ensembles ${\bf
A}(\diamond)$ and ${\bf B}(+)$. A power law $a/t^{\lambda/z}$ with
$\lambda=5.05$ is displayed for comparison.
\\ 
\\

\subsection{Crossover Phenomenon}

It is clear from the last section, that though the asymptotic dynamics is
governed by the new `torque-driven' fixed point, the dynamics at earlier
times $t<t_c$ follows the $g=0$ behavior. This suggests that the dynamics
for arbitrary $g$, may be analyzed as a crossover from the $g=0$ fixed
point characterized by ($z=4$, $\lambda \approx 2$) to the torque-driven
fixed point where ($z=2$, $\lambda \approx 5$).

A simple scaling argument encourages us to think of such a crossover
scenario. On restoring appropriate dimensions, the dynamical equation
Eq.(\ref{eq:dyeq2}) can be rewritten as a continuity equation,
\begin{equation}
\partial{\vec\phi({\bf r},t)}/{\partial t } = - {\bf \nabla }\cdot{\vec j
}
\label{eq:continuity}
\end{equation}
where the ``spin current'' is
\begin{equation}
 \vec{j}_{\alpha} = -\Gamma\left({\bf \nabla }\frac{\delta F[\vec \phi]}{\delta
\phi_{\alpha}} +
\frac{\Omega}{\Gamma}\epsilon_{\alpha\beta\gamma}\phi_{\beta}\nabla\phi_{
\gamma}\right) \, .
\label{eq:current} 
\end{equation}
From a dimensional analysis where we replace $j_{\alpha}$ by the
`velocity' $dL/dt$, we find

\begin{equation} 
\frac{dL}{dt} = \Gamma\frac{\sigma}{L^3} +
\Omega\frac{\sigma M_{0}}{L}\, ,
\label{eq:dimension} 
\end{equation}
where $M_0$, $\sigma$ and $\Gamma^{-1}$ are the equilibrium
magnetization, surface tension and spin mobility respectively.
Beyond a crossover time given by $t_c(g) \sim (\Gamma/M_{0} \Omega)^2 \sim
1/g^2$, simple dimension counting shows that the dynamics crosses over
from $z = 4$ to $z = 2$ in conformity with our numerical simulations. 

\begin{figure}
\centerline{\epsfig{figure=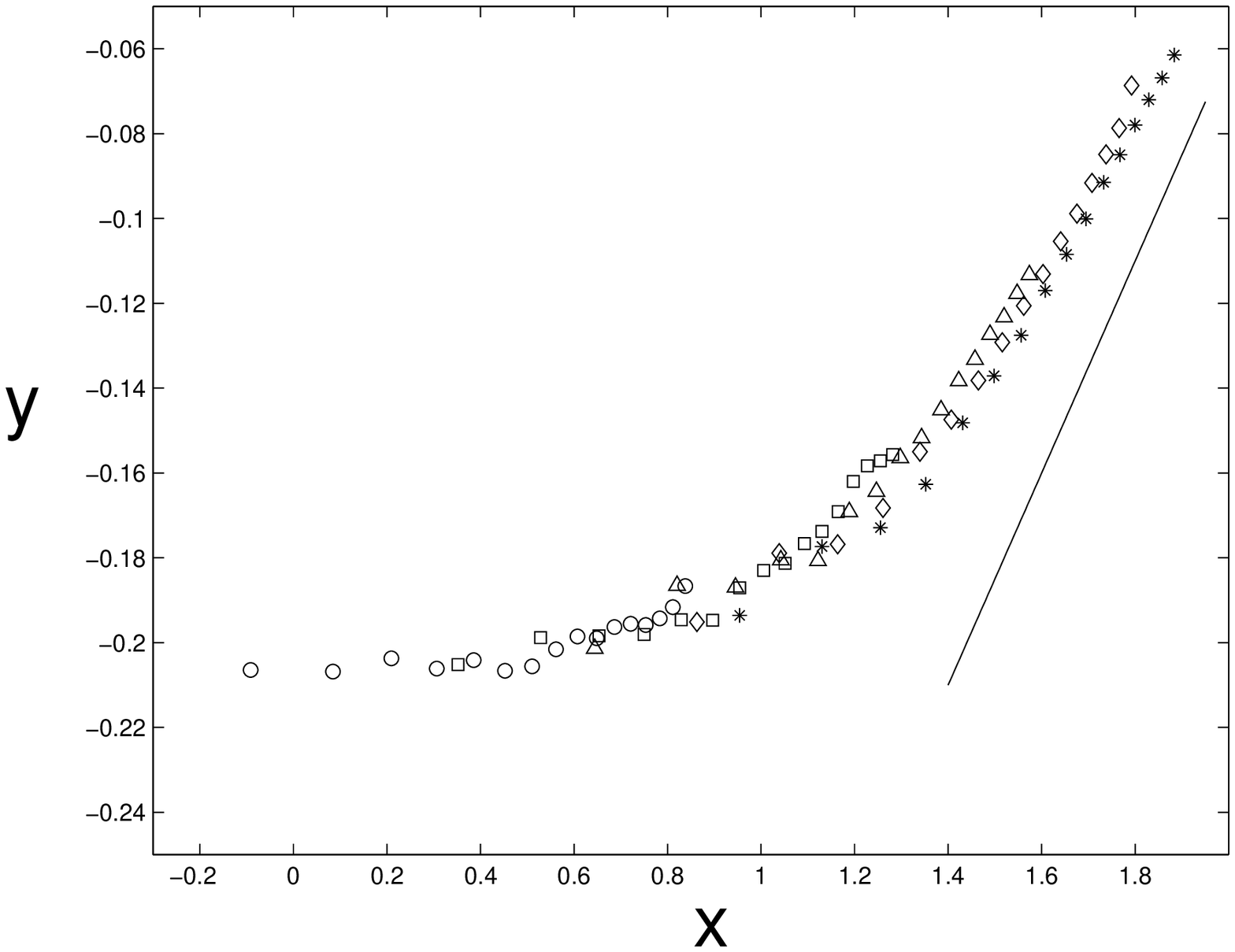,width=7.0cm,height=7.0cm}}
\end{figure} 

FIG. 7. Scaling plot of $y=L(t,g)/t^{1/4}$ versus $x =
tg^{2}$ for $g=0.03(\circ),\,0.05(\Box),\,0.07(\triangle),\,0.09
(\diamond),\,0.10 (\ast)$.  The solid line of slope $0.25$ is the
theoretical estimate of the asymptotic form of the scaling function as
$x\rightarrow \infty$ (see text). \\ 
\\

The crossover physics is best highlighted by numerically demonstrating
crossover scaling of the domain size $L(t,g)$ and the correlation
functions $C(r,t,g)$ and $A(t,g)$.

For instance, Eq.\ (\ref{eq:dimension}) suggests that the domain size
obeys the scaling form $L(t,g) = t^{1/4} s_{m}(tg^{2})$ where the
crossover function $s_m(x)$ is determined from the transcendental
equation, 
\begin{equation}
 x^{1/2} s_{m}(x) - \ln (1+x^{1/2}s_m^2) - 2x = 0\,.
\label{eq:mfcross}
\end{equation}

We shall now argue (and then confirm numerically) that the above scaling
form holds in general. Scaling $r \to r/b$, $t \to t/b^z$ and $g
\to g/b^{y_g}$, scales the domain size by
\begin{equation}
L(t,g) = b\, s(t/b^z,g/b^{y_g})
\label{eq:lscale1}
\end{equation} 
where $y_g$ is the scaling dimension of $g$. We choose $b$ such that
$t/b^z = 1$, which implies
\begin{equation}
L(t,g) = t^{1/z} s(g/t^{y_g/z})\,.
\label{eq:lscal2}
\end{equation}
Setting $g=0$ gives $L(t,g=0) = t^{1/z} s(0)$, telling us that
$z=4$. Thus the scaling form Eq.\ (\ref{eq:lscal2}) is governed by
the $g=0$ fixed point. We therefore need to evaluate $y_g$ at this $g=0$
fixed point. 
We determine $y_g$ by noting the $g$ contribution to
Eq.\ (\ref{eq:dyeq2})
\begin{eqnarray}
\frac{d\vec \phi}{dt} &  \sim   &  g \vec{\phi}\times \delta
                                 F[\vec{\phi}]/\delta \vec{\phi} \nonumber
\\ 
                     &    =    &  g \vec{\phi}\times \vec{\mu}  \nonumber
\\
                     &  \sim   &  g/L^2
\end{eqnarray}
where the last relation is obtained by demanding local
equilibrium (Gibbs-Thomson) on the chemical potential $\vec{\mu}$.
Thus equating dimensions, $[g] = [t^{-1}][L^2] = [L^{-z+2}]=[L^{-2}]$ 
leading to $y_g=-2$. The crossover scaling form for the domain
size can now be read out from Eq.\ (\ref{eq:lscal2}),
\begin{equation}
L(t,g) =  t^{1/4} s(g^{2}t)\,.
\label{eq:lscalfin}
\end{equation}
The $x\to \infty$ asymptote of $s(x)$ can be obtained by demanding that
we recover the `torque-driven' fixed point behavior, which forces
$s(x\to \infty) \sim x^{1/4}$.

We will now check whether this crossover scaling form is seen
in our Langevin simulation.  If the above proposal is true, then the data
should collapse onto the scaling curve $s(x)$ when plotted as
$L(t,g)/t^{1/4}$ versus $tg^{2}$.  Figure 7 shows the results of the
numerical simulation --- the data collapse is not good away from the
asymptotic regimes. To see a better data collapse away from either
fixed point, it is necessary to include corrections to scaling. 

Corrections to scaling come from two sources --- (i) finite time effects
and (ii) nonlinear corrections to the scaling fields \cite{WEG}. Finite
time corrections can be incorporated by introducing finite-time shift
factors $t \to t-t_0$, which can be neglected in the $t \to \infty$ limit. 
Nonlinear corrections to scaling are incorporated by constructing a
nonlinear, analytic function ${\tilde g}(g)$ of the physical fields $g$,
such that it reduces to $g$ in the limit $g\to 0$. The simplest choice of
such a function is
\begin{equation}
\tilde{g}(g) = \frac{g+cg^2}{1+cg^2}\,,
\label{eq:nonlinear}
\end{equation}
leading to a nonlinear scaling variable
\begin{equation}
{\tilde x} = ({\tilde g}(g))^2 (t-t_0)\,.
\label{eq:nonlin}
\end{equation}
The data plotted with respect to this nonlinear scaling variable shows a
much better collapse (Fig.\ 8) when $c$ is chosen to be around $-1.5$ (in
the Figs.\ 8 - 10, the finite time shift $t_0$ was taken to be $0$). The
simple mean-field estimate $s_m({\tilde x})$ plotted for comparison (
Eq.\ (\ref{eq:mfcross})), is exact only at the asymptotes.

\begin{figure}
\centerline{\epsfig{figure=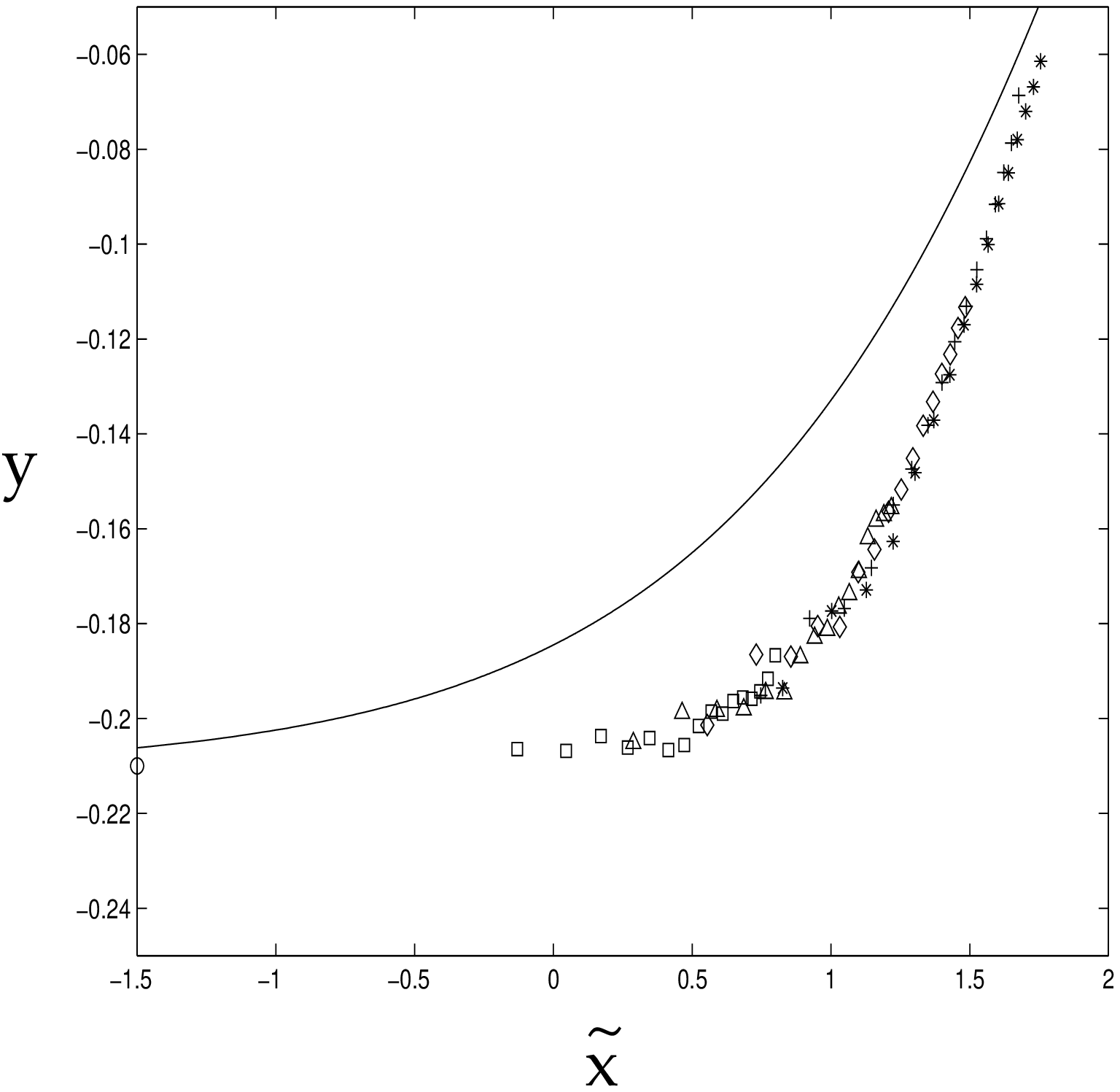,width=8.0cm,height=7.0cm}}
\end{figure} 

FIG. 8. Plot of $y=L(t,g)/(t-t_0)^{1/4}$ versus 
$\tilde x$ when $c\approx-1.5$. The point $\circ$ on the $y$ axis,
represents the value of $y$ as $\tilde{x} \rightarrow 0$.
\\
\\
We have seen in the last section that the equal time
correlation function
$C(r,t,g)$ is unaltered when scaled with the domain size $L$, and so
we expect it to have the following scaling behavior
\begin{equation}
C(r,t,g)=f(r/L,t/L^{z},g/L^{y_g})\,,
\label{eq:crossc}
\end{equation}
where $z$ is the dynamical exponent at the $g=0$ fixed point and $y_g$ is
the scaling dimension of $g$. $L$ is the size of the domain, given by
Eq.\ (\ref{eq:lscalfin}). This readily leads to a two variable scaling
form \cite{WORTIS},
\begin{equation}
C(r,t,g) =  f\left(\frac{r}{t^{1/4}},tg^{2}\right)\,,
\label{eq:crosscor}
\end{equation}
with scaling variables $\rho=r/t^{1/4}$ and $x=tg^{2}$. When $x=0$ and $x
\rightarrow \infty$ then $f(\rho,x)=f_{0}(\rho)$ and
$f(\rho,x)=f_{T}(\rho)$ respectively, where $f_0(\rho)$, $f_{T}(\rho)$ are
the asymptotic scaling functions at $g=0$ and $g\neq 0$.  Again in terms
of the nonlinear scaling variables ${\tilde x}$ and ${\tilde \rho} =
r/(t-t_0)^{1/4}$, we find a very good collapse of the data for
$c\approx-1.2$ (Fig.\ 9).

\begin{figure}
\centerline{\epsfig{figure=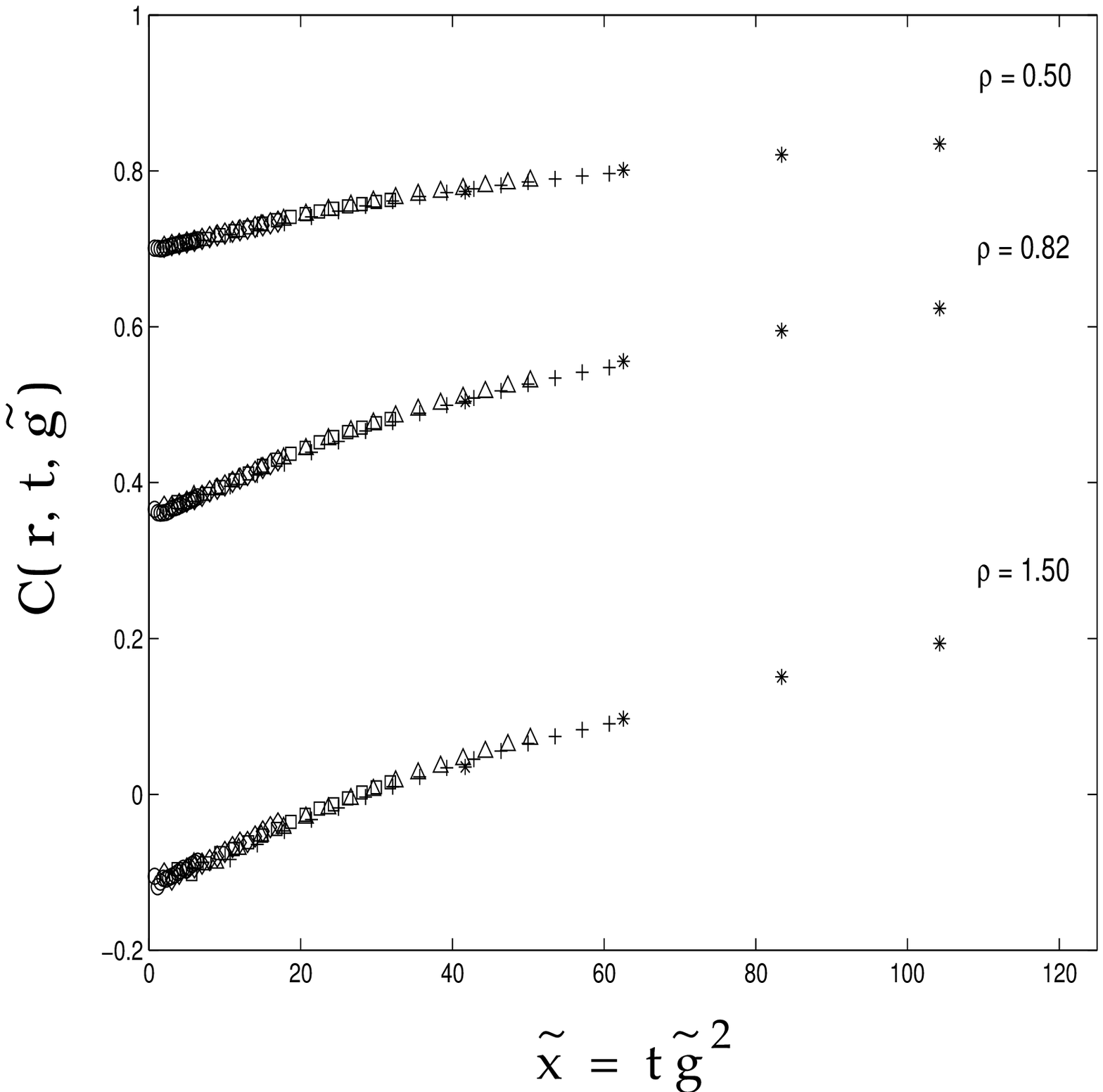,width=8.0cm,height=8.0cm}}
\end{figure} 

FIG. 9. $C(r,t,g)$ versus ${\tilde x}$ at $\tilde \rho
= 0.50,\,0.82,\,$and $1.50$ for $g = 0.03(\circ),\,0.05(\diamond), \,
0.07(\Box),\,0.09(\triangle),\,0.1(+),\,0.3(\ast)$
showing data collapse for $c\approx -1.2$.
\\ 
\\
Similar arguments suggest that the autocorrelation function satisfies the
scaling form 
\begin{equation} 
A(t,g)=t^{-\lambda_0/4}\,a(tg^{2})\,,
\label{eq:crossa} 
\end{equation} 
where $a(x=0)=a_0$ is a constant, and $\lambda_{0}\approx 2.2$ is the
value of the autocorrelation exponent at $g=0$. As $x\to \infty$, the
scaling function $a(x)$ should asymptote to $a(x) \sim
x^{\lambda_{0}/4-\lambda_{T}/2}$, where $\lambda_{T}\approx 5.05$ is the
exponent at the `torque-driven' fixed point. This expectation is borne out
by the numerical simulation (Fig.\ 10), where we have again used the
nonlinear scaling variable ${\tilde x}$ for better collapse.

\begin{figure}
\centerline{\epsfig{figure=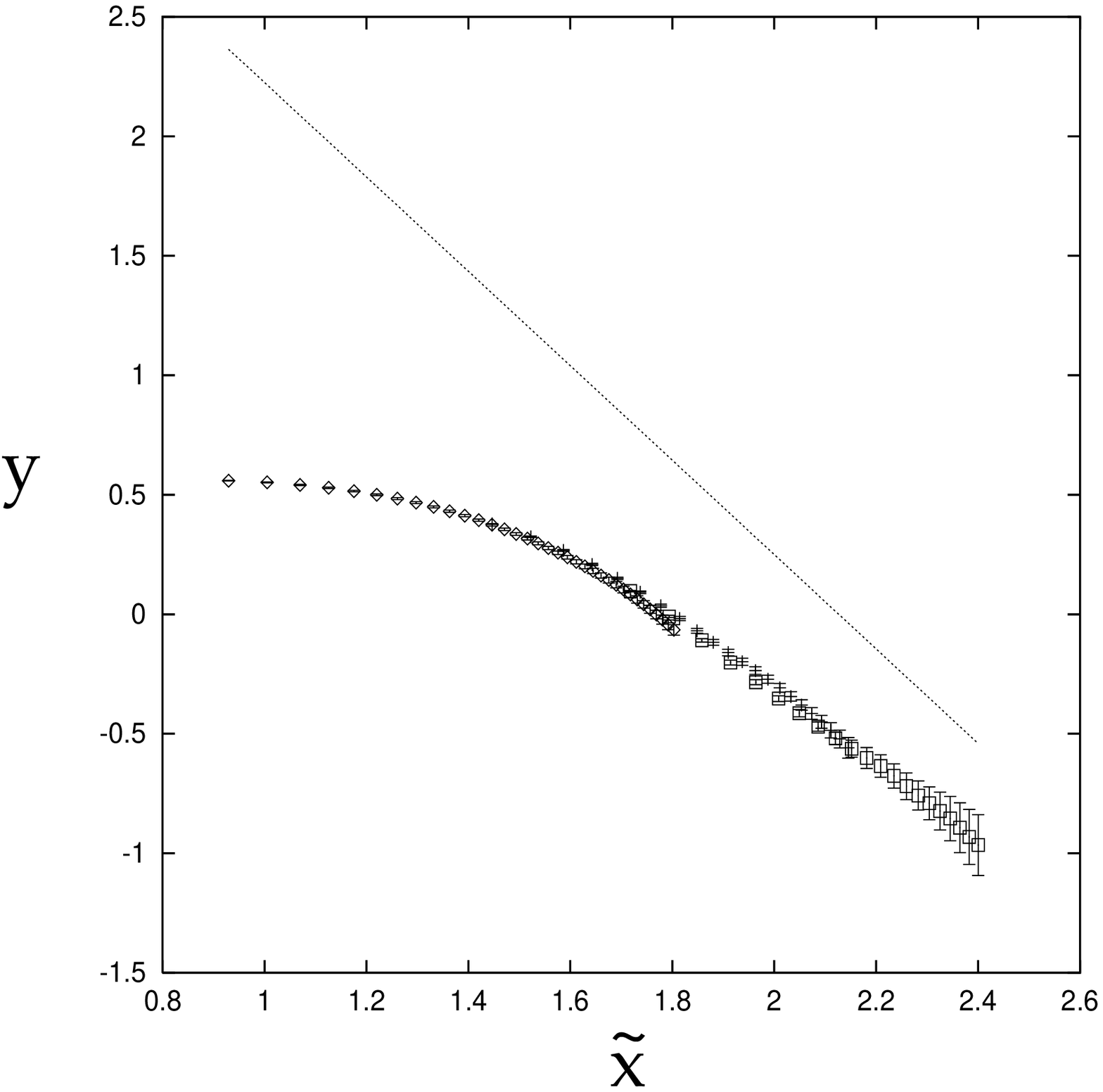,width=8.0cm,height=8.0cm}}
\end{figure} 

FIG. 10. Log-Log plot of $y=A(t,\tilde{g})/t^{-\lambda_0/4}$ versus
${\tilde x}$ for $g = 0.1(\diamond),\,0.2(+),\,0.3(\Box)$
showing data collapse for $c\approx-1.1$. The scaling function
aysmptotes to a line of slope $\lambda_{0}/4-\lambda_{T}/2=-1.95$
as $ {\tilde x} \rightarrow \infty$.
\\
\\
The above discussion clearly indicates that for times $t\ll t_c(g)\sim
1/g^2$, the dynamics is affected by the $g=0$ fixed point while for
$t\gg t_c(g) \sim 1/g^2$, it follows the `torque-driven' fixed point.
Our scaling analysis suggests the following
renormalization group flow diagram,
\\
\\
\begin{figure}
\centerline{\epsfig{figure=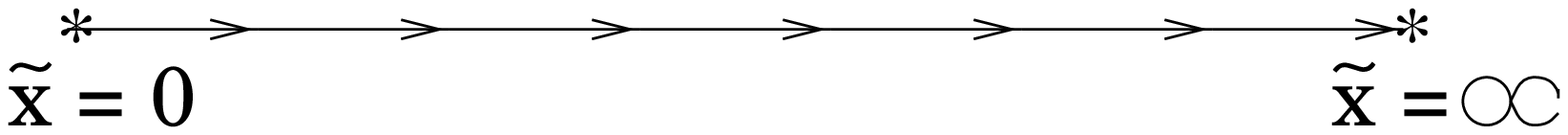,width=8.0cm,height=1.0cm}}
\end{figure} 

\subsection{Failure of Mazenko Closure Scheme\,: Interaction of Defects
with Spin Waves}

We would like to know if the crossover phenomenon described in the last
section can be understood from certain approximate theories of phase
ordering of conserved vector order parameters. In particular could we use
such theories to calculate the crossover scaling functions and the
correlation functions at the `torque-driven' fixed point. The Gaussian
Closure Scheme introduced by Mazenko \cite{MAZENKO1} has been considered
a very successful theory to compute scaling functions of conserved vector
order parameters, and it is to this we turn our attention.

The method consists of trading the order parameter
${\vec \phi} ({\bf r},t)$ which is singular at defect sites, for an
everywhere smooth field ${\vec m}({\bf r},t)$, defined by a nonlinear
transformation,
\begin{equation}
{\vec \phi}({\bf r}, t) = {\vec \sigma}\left({\vec m}({\bf r} ,t)\right)\,.
\label{eq:trans}
\end{equation}
The choice for the nonlinear function ${\vec \sigma}$ is dictated by the
expectation that at late times, the magnitude of ${\vec \phi}$ saturates
to its equilibrium value almost everywhere except near the defect cores.
This suggests that the appropriate choice for ${\vec \sigma}$ is an
equilibrium defect profile,
\begin{equation}
\frac{1}{2}\nabla_{m}^2 {\vec \sigma}\left({\vec m} ({\bf r} ,t)\right) = 
V^{\prime}\left({\vec \sigma}({\vec m }({\bf r} ,t))\right)\,,
\label{eq:defect}
\end{equation}
where $V^{\prime}({\vec x}) \equiv - {\vec x} + ({\vec x} \cdot {\vec
x})\,{\vec
x}$. The auxiliary field ${\vec m}$ now has the natural interpretation as
the position vector from the nearest defect core.  Implicit in this choice
is that smooth configurations such as spin waves relax fast and so
decouple from defects at late times. The simplest nontrivial solution of
Eq.\ (\ref{eq:defect}) is the hedgehog configuration,
\begin{equation}
{\vec \sigma}\left({\vec m}({\bf r},t)\right) = \frac{{\vec m}({\bf r}, t 
)} {\vert{\vec m}({\bf r},t) \vert}\,g({\vert{\vec m}\vert})\,,
\label{eq:hedgehog}
\end{equation}
where $g(0)=0$ and $g(\infty)=1$.  

Equation (\ref{eq:dyeq2}) can be
used to derive an equation for the correlation function $C(12) \equiv
\langle {\vec \phi}({\bf r}_1, t_1) \cdot {\vec \phi}({\bf r}_2,
t_2)\rangle$. Substituting for ${\vec \phi}$ (Eqs.\ (\ref{eq:trans}), 
(\ref{eq:hedgehog})) in the right hand side of the
resulting equation, we get
\begin{eqnarray}
{\partial_t}{C(12)}   &  =  &
                            -\nabla_1^2 \left[
                            \nabla_1^2 C(12) - \langle\, {\vec \sigma}
                            ({\vec m}(2)) \cdot V'({\vec \sigma}
                            ({\vec m}(1)))\,\rangle \right] \nonumber \\
                      &     &
                            +\, g \,\langle\, {\vec \sigma}({\vec m}(2)) 
                           \cdot {\vec \sigma}({\vec m}(1)) \times 
                           \nabla_1^2 {\vec \sigma}({\vec m}(1))\,\rangle\,\,. 
\label{eq:langm}
\end{eqnarray}
The Gaussian Closure Scheme assumes
that each component of ${\vec m}({\bf r},t)$ is an independent gaussian
field with zero mean at all times. This implies that the joint probability
distribution $P(12) \equiv P({\vec m}(1), {\vec m}(2))$ is a product of
separate distributions for each component and is given by  \cite{BRAY},

\begin{eqnarray}
\prod_{\alpha} {\cal N} \exp
      \bigg\{\, - \frac {1} {2\left(1-\gamma^{2}\right)} \bigg(
     \frac{m_{\alpha}^2(1)}{S_0(1)}+\frac{m_{\alpha}^2(2)}{S_0(2)}
	\nonumber \\
- \frac{2 \gamma m_{\alpha}(1)m_{\alpha}(2)} 
        {\sqrt{S_0(1)S_0(2)}} \bigg) \bigg\}\,,
\label{eq:mazprob} 
\end{eqnarray}
where $$ {\cal N} = \frac{1}{2\pi\sqrt{(1-\gamma^2)S_0(1)S_0(2)}} $$
and 
\begin{equation}
\gamma \equiv \gamma(12) = \frac {C_0(12)}{\sqrt{S_0(1)S_0(2)}}\,.
\label{eq:gamma}
\end{equation}
The joint distribution has been written in terms of the second moments
$S_0(1) = \langle m_{\alpha}(1)^2 \rangle$
and $C_0(12) = \langle m_{\alpha}(1)
m_{\alpha}(2) \rangle$.

With this assumption, the right hand side of Eq.\ (\ref{eq:langm}) 
simplifies to,
\begin{eqnarray}
\frac{\partial C(12)}{\partial t_1} & = & -\nabla^2 \left[ \nabla^2 C(12)
                   + \frac{\gamma}{2S_0(1)}\frac{\partial
                     C(12)}{\partial \gamma } \right] \nonumber \\
             &  &  +\,g\, \langle \,{\vec \sigma}({\vec m}(2)) 
                     \cdot {\vec \sigma}({\vec m}(1))
	 	 \times \nabla^2 {\vec \sigma}({\vec m}(1))\,\rangle\,,
\label{eq:mazcorr}
\end{eqnarray}
where the laplacian is taken with respect to ${\bf r}_1$.
With the joint probability distribution given by Eq.\
(\ref{eq:mazprob}),
it is clear that the last term in the above equation vanishes,
implying that the torque is irrelevant at late times. This result of the
Gaussian Closure Scheme, is in direct contradiction to the
results of the last two sections.

What has gone wrong ? There are two possible sources of error
\begin{enumerate}

\item The Gaussian assumption for the probability distribution of $\vec m$
is invalid.

We show below that while the Gaussian assumption leads to an internal
inconsistency, it may be remedied by considering corrections to the
gaussian distribution. This however does not solve the above
contradiction.

\item The order parameter $\vec \phi$ cannot be written in terms of the
defect field $\vec m$ alone.

\end{enumerate}

We will first question the Gaussian assumption, on the lines suggested by
Yeung et. al. \cite{CHUCK} in the case of a conserved scalar (Ising) order
parameter. We will do this for the case when $g=0$, the $g\neq0$
analysis follows similarly. 

The equal time correlation function may be derived from Eqs.\
(\ref{eq:mazcorr}), (\ref{eq:mazprob}) and takes the form
\cite{MAZENKO1}

\begin{equation}
C(r,t)=\frac{3\gamma}{2\pi}\left[B\left(2,\frac{1}{2}\right)\right]
^{2} F\left(\frac{1}{2},\frac{1}{2},\frac{5}{2};\gamma^2\right) 
\label{eq:hyper}
\end{equation}
where $B(x,y)$ and $F(a,b,c;z)$ are the Beta and hypergeometric
functions respectively and $\gamma$ is given in Eq.\ (\ref{eq:gamma}) . 
We may expand the hypergeometric function as a power series in $\gamma$ 
\cite{ABRAM} and then take its fourier transform,
\begin{eqnarray}
S({\bf k},t) &=& \sum_{p=0}^{\infty}\,\int d{\bf k}_1 \ldots d{\bf
k}_{2p+1} \, \bigg[
a_{p}\gamma_{{\bf k}_1}(t)\gamma_{{\bf k}_2}(t) \ldots
\gamma_{{\bf k}_{2p+1}}(t) \nonumber \\
&  & \delta({\bf k}+{\bf k}_1+\ldots+{\bf k}_{2p+1})\bigg]
\label{eq:hyperk}
\end{eqnarray}
where the spectral density $\gamma_{\bf k}$ is the fourier transform of
$\gamma(r,t)$ and the expansion coefficients,
\begin{equation}
a_p=\frac{9}{8\pi^{3/2}}\frac{\left[\Gamma(p+1/2)\right]^2}{\Gamma(p+5/2)p!}
\, \left[ B\left(2,\frac{1}{2}\right)\right]^{2}\,,
\label{eq:coeff}
\end{equation}
are strictly positive for $p\geq0$. If Eq.(\ref{eq:hyperk}) has to satisfy
the conservation law $S(k=0,t)=0$, it is clear that
$\gamma_{\bf k}(t)$ should be $negative$ at some values of ${\bf k}$. This
is inconsistent with the definition Eq.\ (\ref{eq:gamma}) which implies
$\gamma_{\bf k}(t) \geq 0$ for all ${\bf k}$. This definition is a
consequence of the Gaussian approximation.

\begin{figure}
\centerline{\epsfig{figure=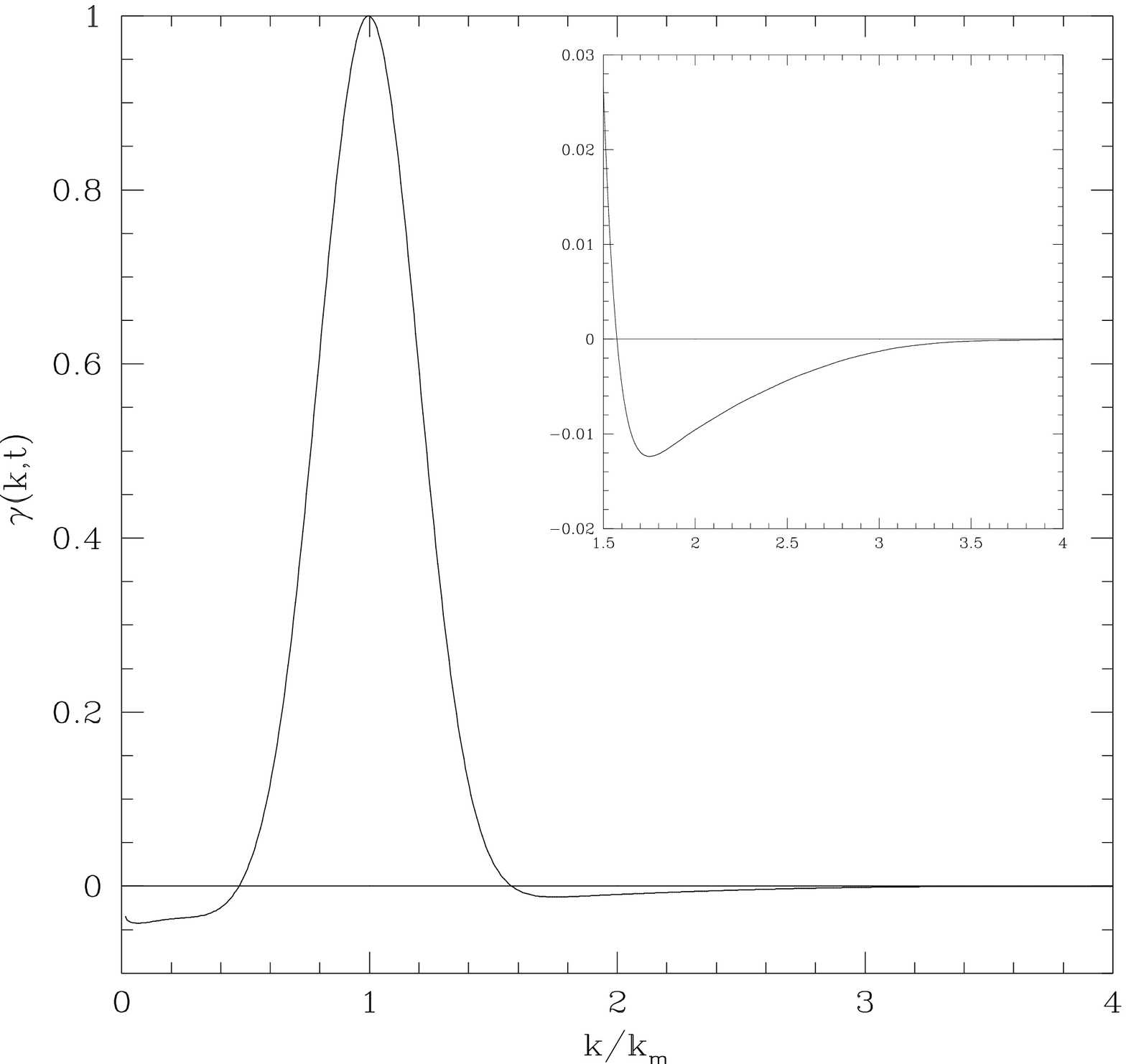,width=7.0cm,height=7.0cm}}
\end{figure} 

FIG. 11. The spectral density $\gamma(k,t)$ at $t=3600$
becomes negative for $0\leq k/k_{m}<0.5$ and for $1.5<k/k_{m}<3.0$
(inset).

\vskip.3cm

To determine the range of values of ${\bf k}$ for which $\gamma_{\bf k}$
is negative, we numerically evaluate the fourier transform of
$\gamma(r,t)$ after inverting Eq.\ (\ref{eq:hyper}). This is prone to
numerical errors because of statistical errors in our computed $C(r,t)$.
For instance, a numerical integration of $\int d{\bf r}\,C(r,t)$ gives
a nonzero value whereas it should be identically zero because of the
conservation law. This is reflected in large errors in $\gamma({\bf k},t)$
at small ${\bf k}$.  We therefore adopt the following procedure.  We fit a
function $C_{f}(x)$ to the equal time correlation function
$C(r,t)$ and use this to extract $\gamma({\bf k},t)$ from the
Eq.\ (\ref{eq:hyper}). The fitting function has been taken to be
\begin{equation}
C_{f}(x)=\frac{\sin(x/L)}{(x/L)}
\left[1+a\left(\frac{x}{L}\right)^2\right]\,\exp[-b(x/L)^2]
\label{eq:fitfunc}
\end{equation}
which is similar to the analytic form given in Ref.\ \cite{FIT}. Note
that only $b$ and $L$ are independent fitting parameters, $a$ is
determined from the condition $S_{f}(k=0)=0$. This function with
$L=1.5106\pm 1.01\times 10^{-4}$ and $b=0.0202\pm 2.14\times 10^{-4}$
gives a very good fit to $C(r,t)$ upto the fourth zero of the function.  
We observe (Fig.\,11) that the spectral density, which should be a
strictly positive function of its arguments, becomes negative for $k/k_{m}
< 0.5 $ ($\gamma(k,t)$ is peaked at $k_{m}$)  and in the range
$1.5<k/k_{m}<3.0$.

Our demonstration suggests that a purely gaussian theory for the
distribution of ${\vec m}$ is internally inconsistent. This may however be
remedied by considering corrections to the purely gaussian distribution,
as suggested by Mazenko \cite{MAZENKO2} for the scalar (Ising) order
parameter. 

In order to help us understand the nature of the corrections, let us first
numerically evaluate the probability distribution of $\vec m$. We
determine $\vec{m}$ by choosing $g(\vert \vec{m}\vert)$ in such a way as
to make Eq.\ (\ref{eq:hedgehog}) invertible. A convenient choice is
\begin{equation}
\vec{\phi}=\vec{\sigma}(\vec{m})=\frac{\vec{m}}{\sqrt{1+|\vec{m}|^2}} \, .
\label{eq:sigmoid}
\end{equation} 
We now compute the asymptotic single point probability density $P(m_1({\bf
r},t))$ on a $50^3$ lattice averaged over $18$ initial configurations for
$g=0,\,0.3,\,0.4$ and $0.5$.  The probability density obeys a scaling form
at late times (Figs.\ 12 and 13), $P(m_1,t) = P(m_1/L(t))$, where the
length scale $L(t)=\sqrt{\langle m_{1}^2 \rangle} \sim t^{1/z}$.  
Moreover Fig.\ 14 shows that the scaled distribution of $\vec m$ is
identical for $g=0$ and $g\neq0$ (the joint probability distributions are
however very different).  It is clear from Figs.\ 12 - 14, that the
asymptotic distributions show marked deviations from a simple gaussian. To
highlight these deviations, we plot the scaled $\log(-\log(P(m_1)))$
versus $\log(m_1^2)$ (Fig.\ 15), a gaussian distribution would have given
a straight line with slope $-1$.

\begin{figure}
\centerline{\epsfig{figure=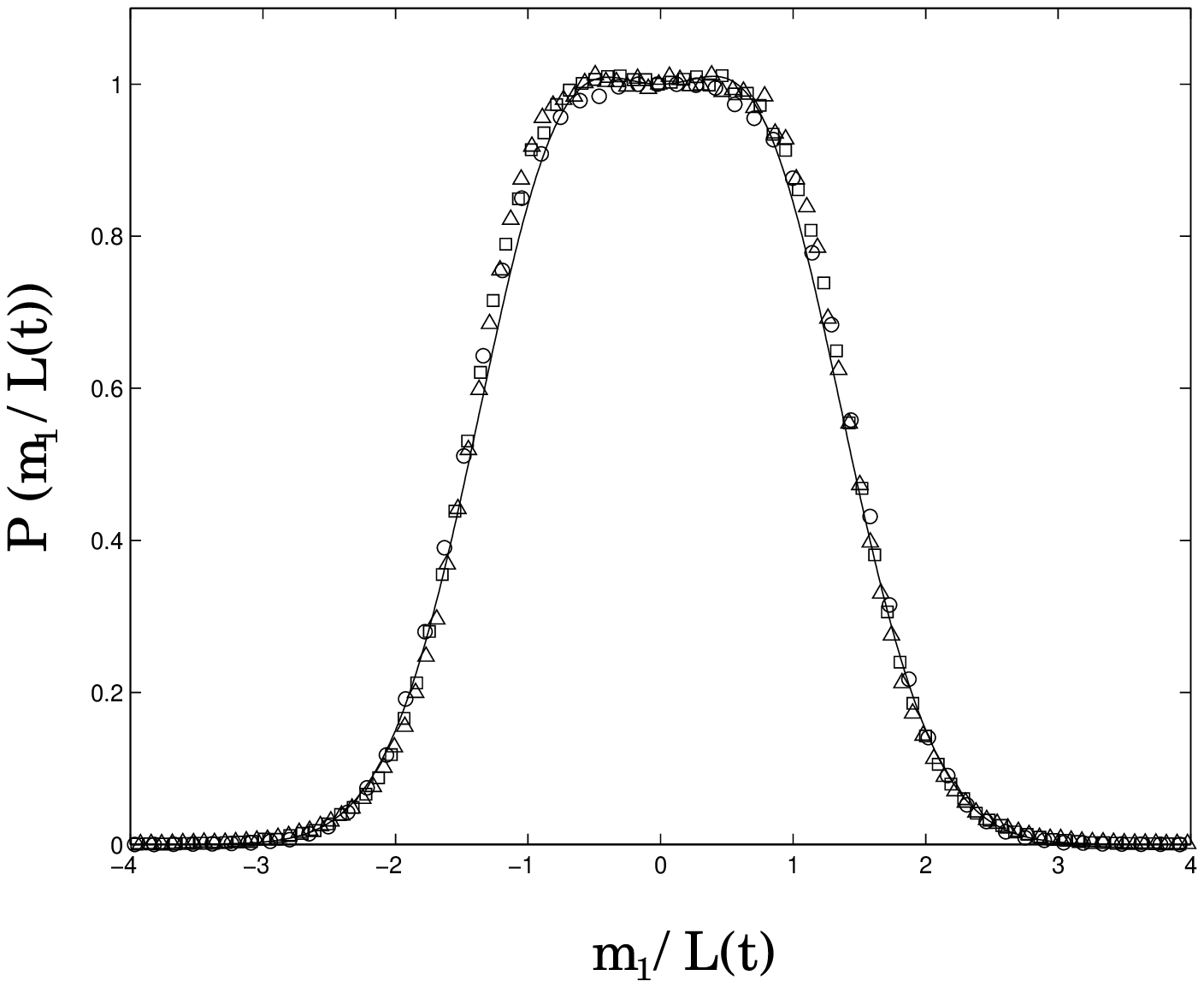,width=7.0cm,height=7.0cm}}
\end{figure}
 
FIG. 12. Scaling plot of the un-normalized $P(x=m_1/L(t))$ for $g=0$ at
different times $t = 900(\circ),\,3600(\Box),\,6300(\triangle)$. Solid
line is a fit to Eq.\ (\ref{eq:hermite}).
\\
\\
\begin{figure}
\centerline{\epsfig{figure=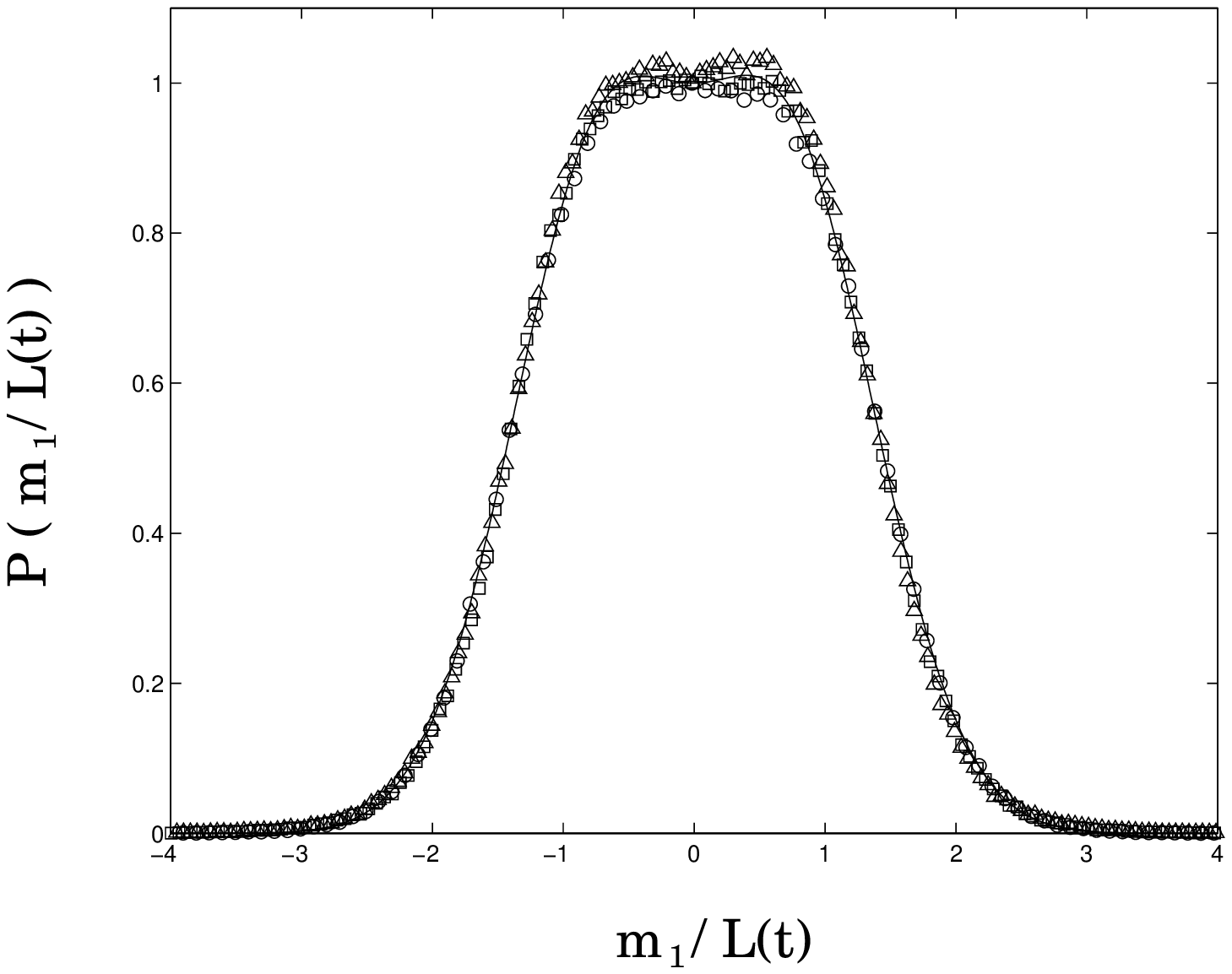,width=7.0cm,height=7.0cm}}
\end{figure}

FIG. 13. Scaling plot of the un-normalized $P(x=m_1/L(t))$ for $g=0.3$ at
different times $t = 1350(\diamond),\,3600(+),\,5400(\Box)$. Solid line is
a fit to Eq.\ (\ref{eq:hermite}).
\\
\\
\begin{figure}
\centerline{\epsfig{figure=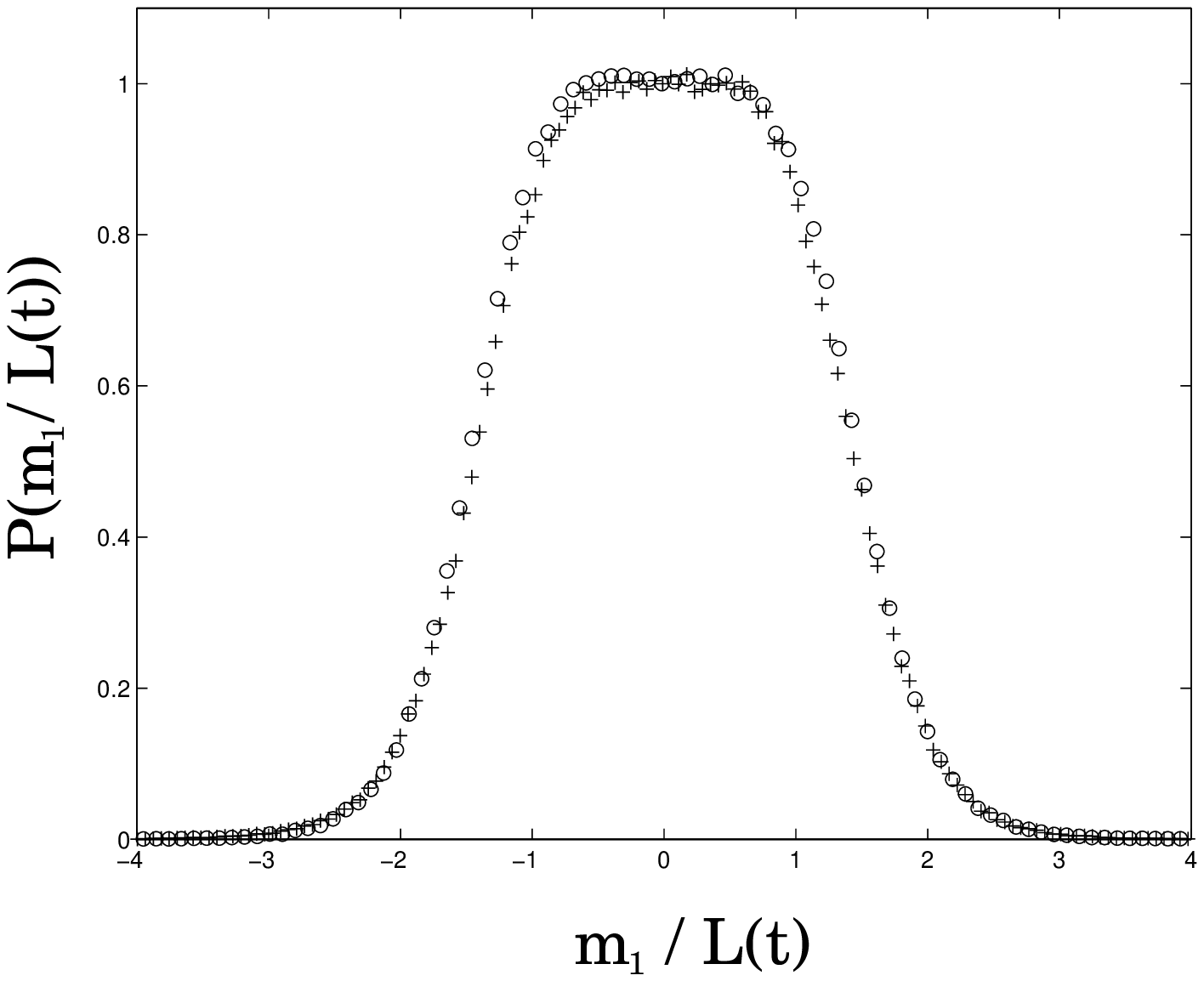,width=8.0cm,height=7.0cm}}
\end{figure}
 
FIG. 14. Scaling plot of the un-normalized $P(x=m_1/L(t,g))$ 
for $g=0(\diamond)$ 
and $g=0.3(+)$ at $t=4500$ showing that the distributions are 
identical within error bars.
\\
\\
\begin{figure}
\centerline{\epsfig{figure=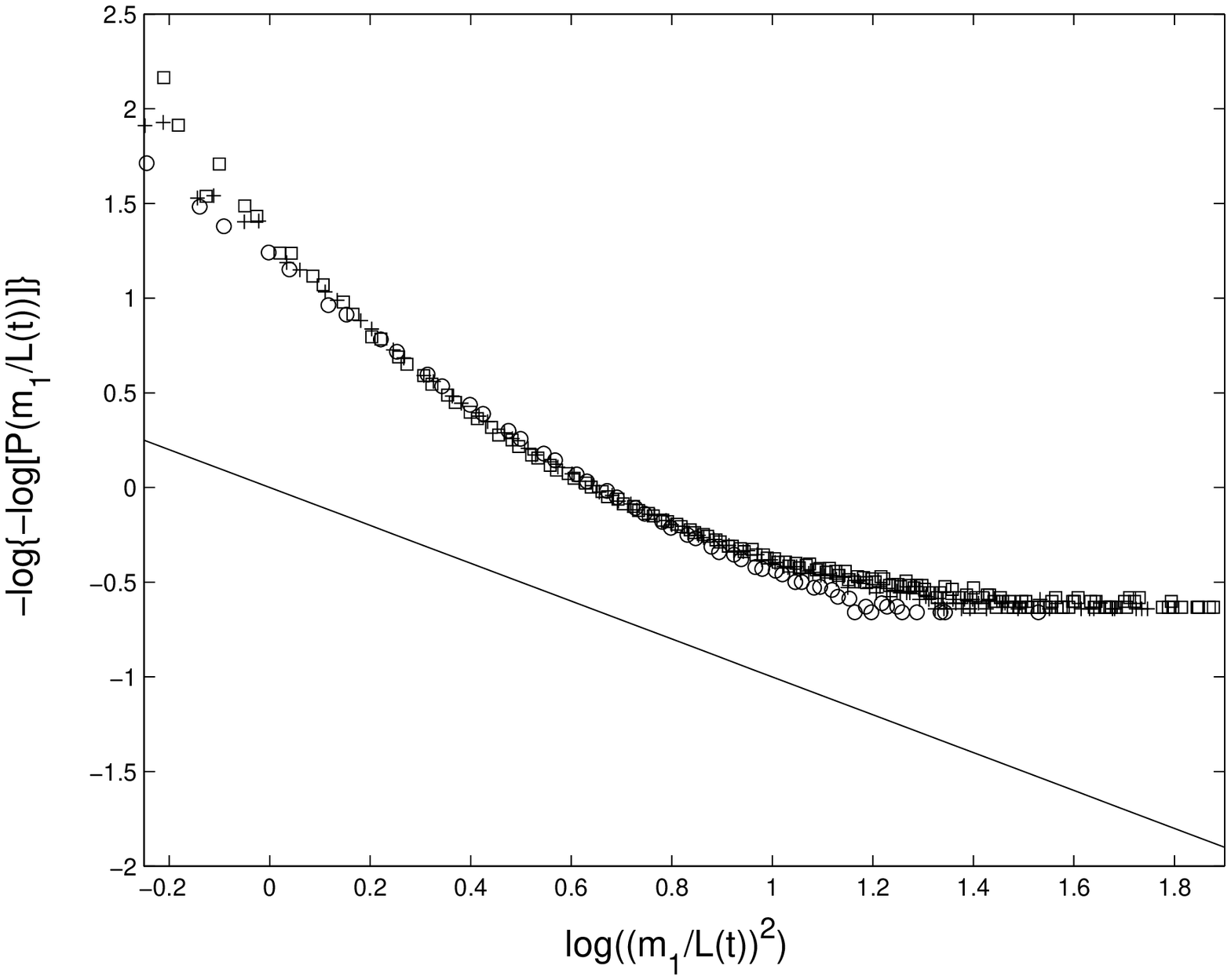,width=8.0cm,height=7.0cm}}
\end{figure} 

FIG. 15. 
Deviation of $P(x=m_1/L)$ from gaussian (straight line) for $g=0$. Data
have been collected at times $t = 900(\diamond),\,3600(+),\,6300(\Box)$.
\\
\\ 
Figures 12 - 14 suggest that the  
deviations from gaussian can be computed by expanding $P(m)$ in a Hermite
polynomial basis $H_n$ (a strategy advocated in Ref.\ \cite{MAZENKO2} for 
the scalar (Ising) dynamics), 
\begin{equation}
P(x) = \sum_{n=0}^{\infty} p_n H_n(x)\, e^{-x^2} \,,
\label{eq:hermite}
\end{equation}
where $x=m_1(r,t)/\sqrt{S_0(r,t)}$ and $H_0(x) = 1$, $H_1(x)=2x$ and
$H_{n+1}(x) = 2x H_{n}(x)-2n H_{n-1}(x)$. The dark line in Figure 12 is an
accurate fit to the $g=0$ data, with $p_0=1$, $p_1=1.33 \times 10^{-3}\pm
6.0\times 10^{-5}$, $p_2=0.2352 \pm 3.8 \times 10^{-5}$, $p_3=1.55 \times
10^{-4} \pm 1.5 \times 10^{-5}$, $p_4=5.542 \times 10^{-3} \pm 7.0 \times
10^{-6}$. Similarly in Fig.\ 13, the dark line is an accurate fit to the
$g=0.3$ data with $p_0=1$, $p_1=3.95 \times 10^{-3} \pm 5.5 \times
10^{-5}$, $p_2=0.2899 \pm 1.3 \times 10^{-5}$, $p_3=5.35 \times 10^{-4}
\pm 1.3 \times 10^{-5}$, $p_4=1.1913 \times 10^{-2} \pm 7.0 \times
10^{-6}$. Indeed the odd coefficients are zero to within numerical
accuracy, indicating that the distribution is even.

It is conceivable that such corrections would be able to salvage the
inconsistency issue, since an additive term to the right hand side of Eq.\
(\ref{eq:hyperk}) would not allow us to assert that $\gamma_k$ should be
negative for some values of $k$.

Though the remedy suggested cures the inconsistency problem, it will still
give a zero value to the torque contribution in Eq.\ (\ref{eq:mazcorr}),
as long as the probability distribution of each component of $\vec m$ is
even and independent. We have already demonstrated that the single point
distribution is even, now we shall show that each cartesian component of
$\vec m$ is independently distributed.

We numerically calculate $P(m_1(1),m_2(2))$ (which we label $P(x,y)$) at
equal times $t_1=t_2=t$ and arbitrary separation, say $|{\bf r}_1-{\bf
r}_2|=4\sqrt 3$ for $g=0.3$ (Fig. \ 16). 

\begin{figure}
\centerline{\epsfig{figure=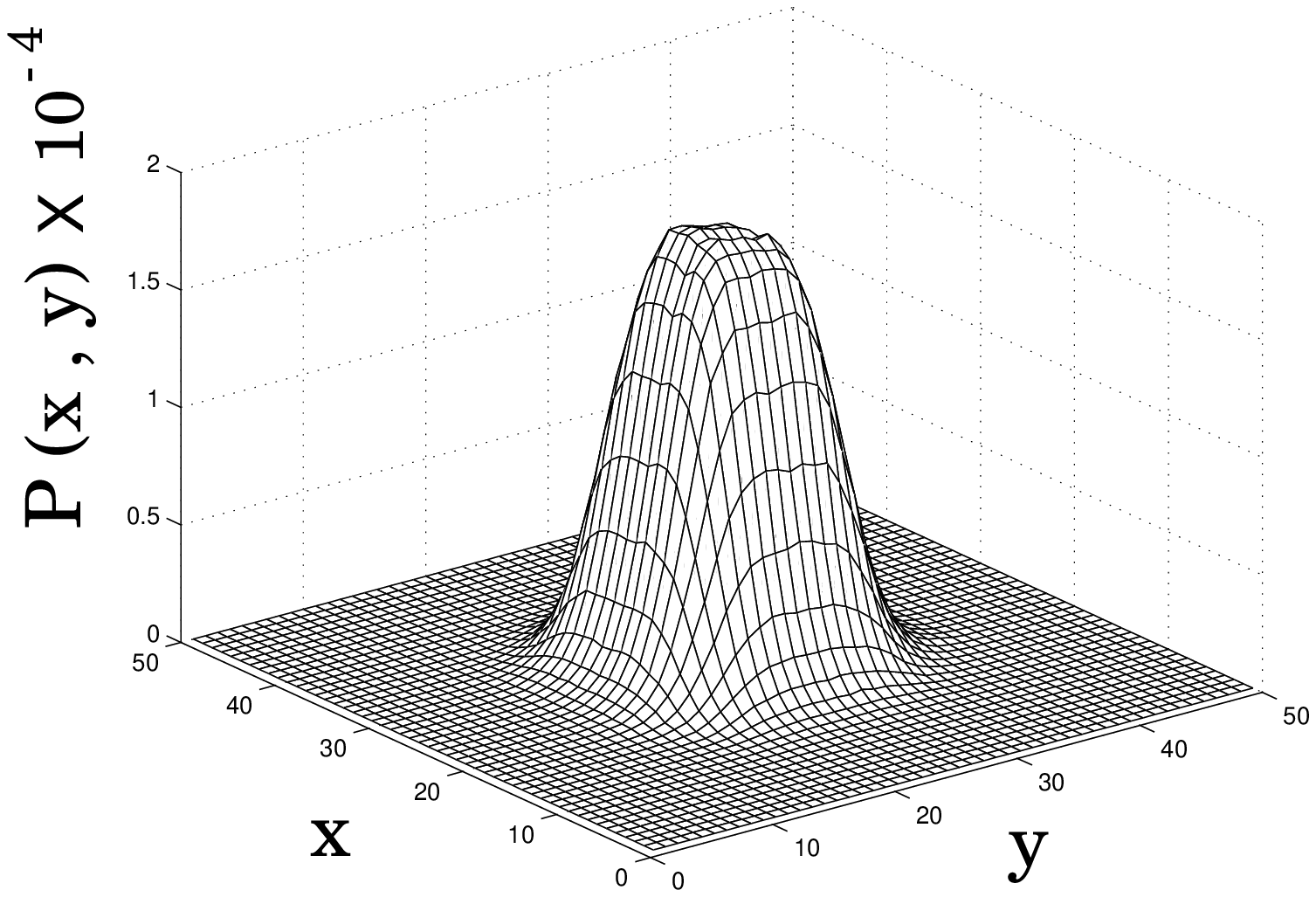,width=7.0cm,height=7.0cm}}
\end{figure} 

FIG. 16. Normalized joint probability distribution $P(x,y)$ where
$x=m_1(1),\, y=m_2(2)$  for $g=0.3$
at $t=2250$ and $|{\bf r}_1-{\bf r}_2| = 4 \sqrt{3}$ 
(averaged over $18$ initial configurations).
\\
\\
To show that the joint distribution is independent in each component, 
we plot the difference
$\Delta(x,y)=P(x,y)- P(x)P(y)$ for $g=0.3$ (Fig.\ 17) 
and find it to be zero within the accuracy of our
numerical computation.

\begin{figure}
\centerline{\epsfig{figure=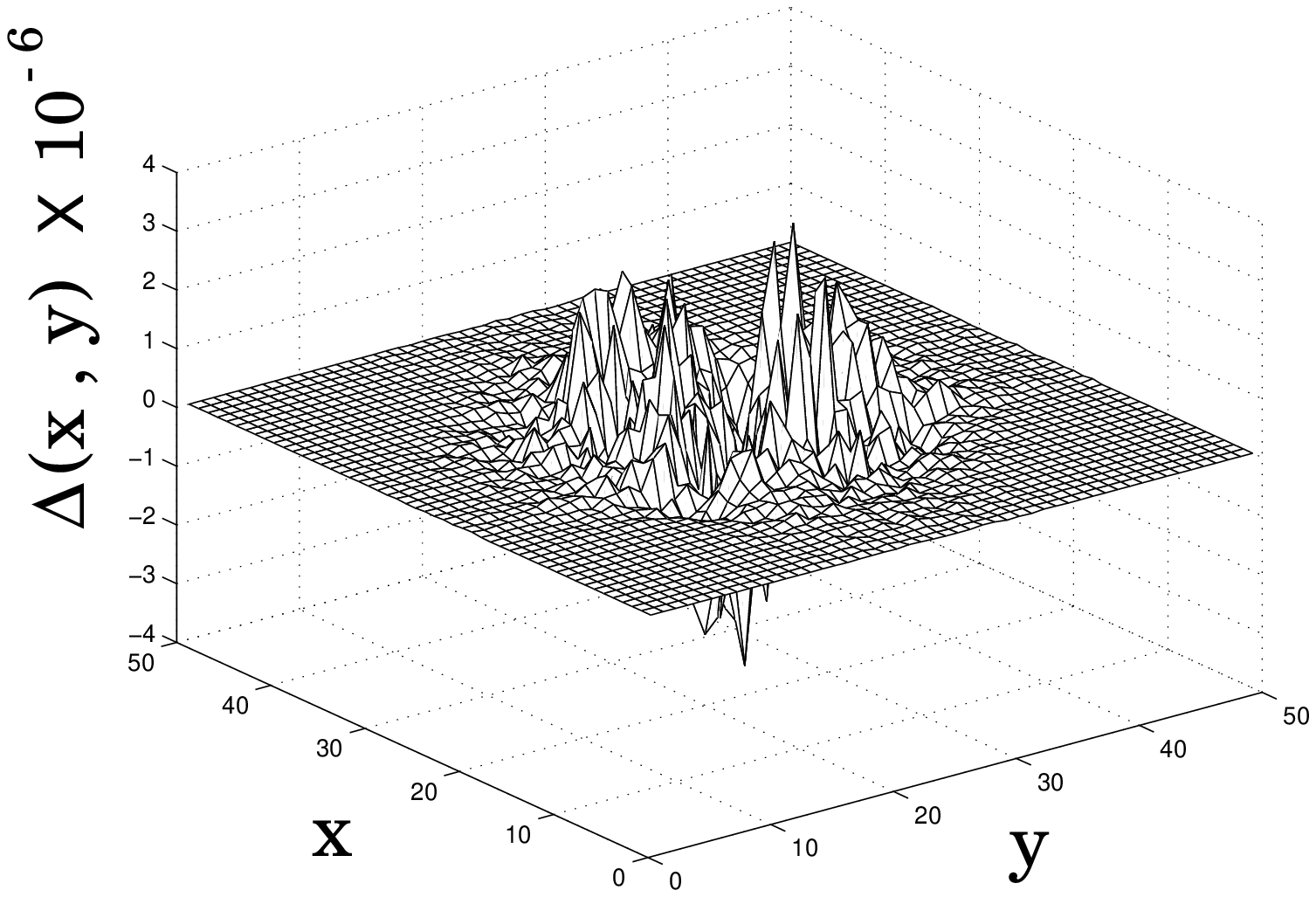,width=7.0cm,height=7.0cm}}
\end{figure} 

FIG. 17. Plot of $\Delta(x,y)$ where
$x=m_1(1),\,  y=m_2(2)$
at $t=2250$ and $|{\bf r}_1-{\bf r}_2| = 4 \sqrt{3}$ for $g=0.3$. 
The maximum magnitude of
$\Delta$ is of the order of errors in $\Delta(x,y)$.
\\
\\
\\
We are thus forced to admit the second possibility, namely that the order
parameter $\vec \phi$ cannot be written in terms of $\vec m$ alone. For in
transforming the spins ${\vec \phi}$ exclusively to ${\vec m}$ we have
implicitly ignored spin waves. A most direct demonstration of this is to
compare $C_{3 \vec{\phi}} = \langle \vec{\phi}(1)\cdot
(\vec{\phi}(2)\times\nabla_{2}^{2}\vec{\phi}(2))\rangle$ with the
defect-only contribution $C_{3 \widehat{m}}=\langle \widehat{m}(1)\cdot
(\widehat{m}(2)\times\nabla_{2}^{2}\widehat{m}(2))\rangle$ (where
$\vec{m}$ is computed by inverting Eq.\ (\ref{eq:sigmoid})).

We find that for $g=0$ both $C_{3 \vec{\phi}}$ and $C_{3 \widehat{m}}$ are
zero within error bars (Fig.\ 18). This is true even at very early times
which implies that in the absence of the torque the spinwaves decay very
fast compared to the relaxation timescale of the defects.  On the other
hand, when $g\neq0$, we find that the two correlators behave very
differently. Figure 19 clearly shows that even at late times, $C_{3
\vec{\phi}}$ is non zero while the defect-only contribution $C_{3
\widehat{m}}$ is zero within errorbars.

This suggests the following decomposition in terms of defect fields
(singular part) and spin-waves (smooth part),
$\vec{\phi}=\vec{\sigma}(\vec m)+\vec{u}$, when $g\neq0$. Such a
decomposition gives rise to contributions to $C_{3 \vec{\phi}}$ reflecting
the interaction between defects and spin-waves.

\begin{figure}
\centerline{\epsfig{figure=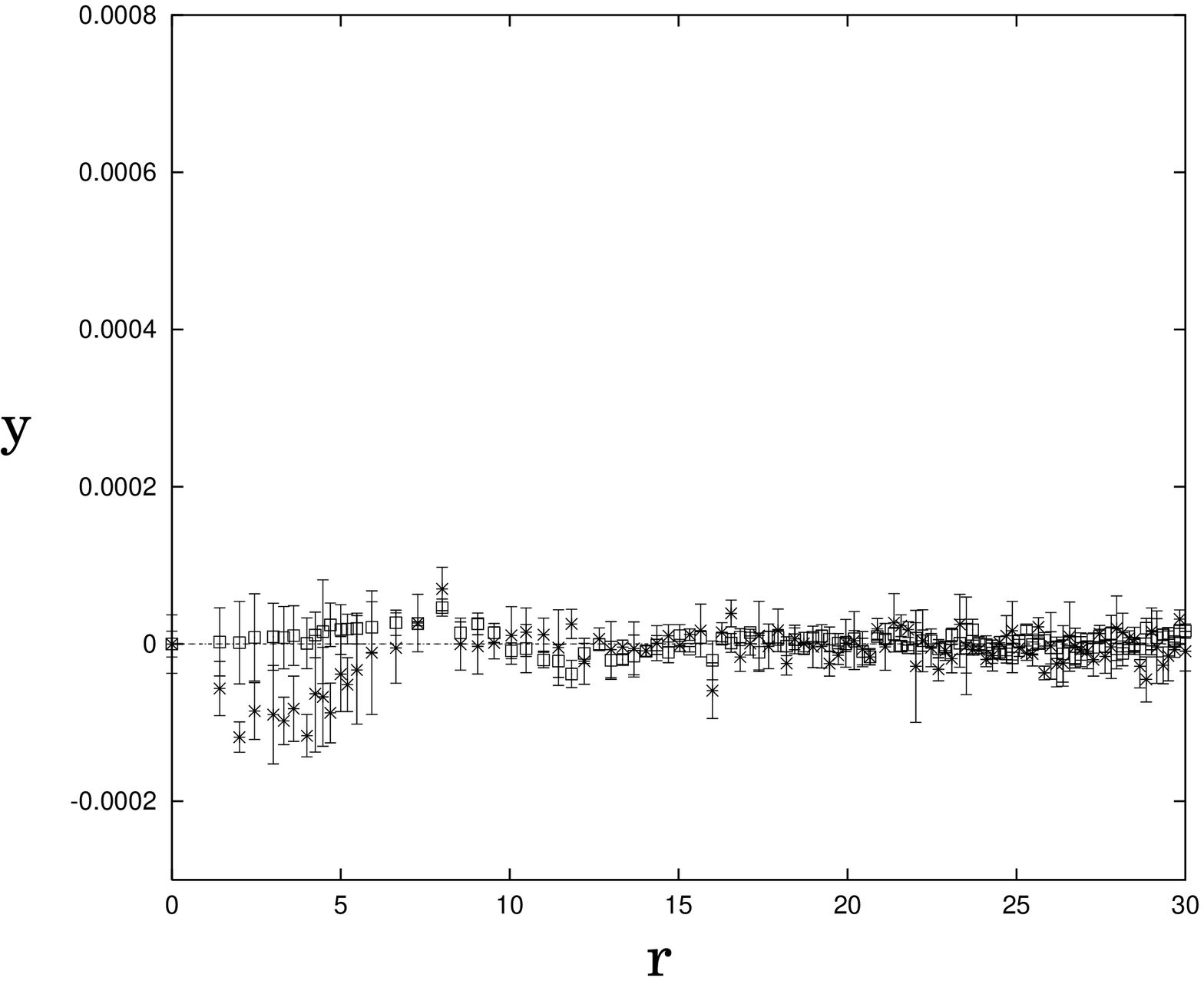,width=7.0cm,height=7.0cm}}
\end{figure} 

FIG. 18. $y=C_{3 \vec{\phi}}(r)$ $(\Box)$ and
$y=C_{3 \widehat{m}}(r) (\ast)$ at $t=3600$ and $r = |{\bf r}_1-{\bf
r}_2|$ for $g=0$ are zero within the error bars (averaged over $5$ 
initial configurations).
\\
\\
\begin{figure}
\centerline{\epsfig{figure=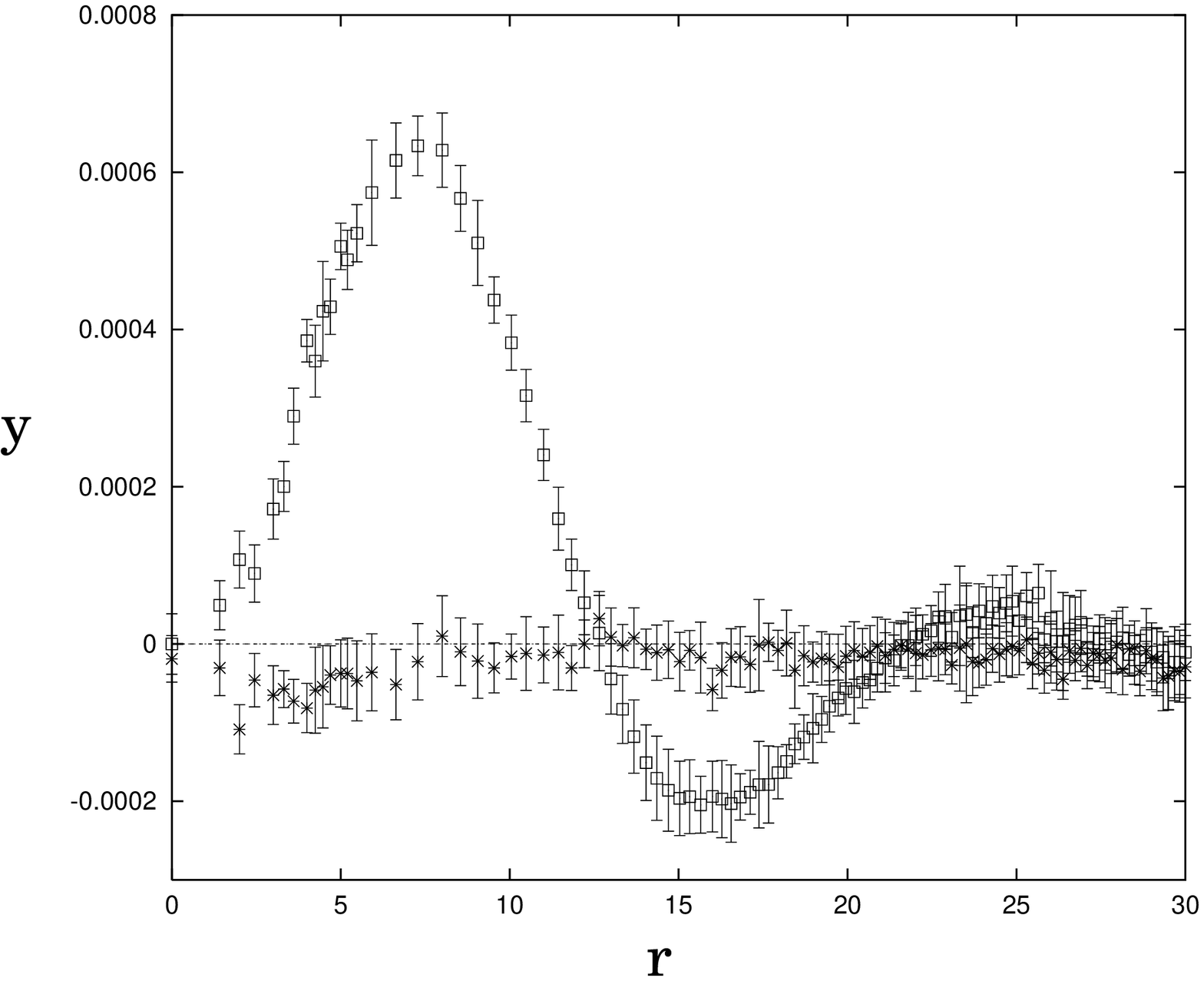,width=7.0cm,height=7.0cm}}
\end{figure} 

FIG. 19. $y=C_{3 \vec{\phi}}(r)$ $(\Box)$ and $y=C_{3 \widehat{m}}(r)
(\ast)$ at $t=3600$ and $r = |{\bf r}_1-{\bf r}_2|$ for $g=0.3$ are
distinctly different (averaged over $5$ initial configurations). $C_{3
\widehat{m}}(r) (+)$, which has contributions from defects alone, is zero
(within errorbars), whereas $C_{3 \vec{\phi}}(r)$, which in addition
involves spin-wave excitations, is non zero.
\\
\\
We conclude this long meandering section by recounting its salient
results. When $g\neq0$, typical spin configurations at late times consist
of slowly moving defects and long-lived spin-waves which interact with
each other. The asymptotic spin distribution cannot be written in terms 
of the distribution of defects alone. When $g=0$, the
spin waves decay faster, leading to a decoupling of the spin waves and
defects at late times. 

\section{Ordering Dynamics at $T = T_c$}
We end this study with a brief discussion of the ordering dynamics Eq.\
(\ref{eq:dye}) of Heisenberg spins quenched to the critical point. The
critical dynamics of this model (called Model J in this context) was
investigated some time ago by Ma and Mazenko \cite{MA}. On the other hand,
the dynamical renormalization group formalism for quench dynamics set up
by Janssen $et. al.$, has been used to study Models A - C
\cite{JANS,KISS}. In this section we use the dynamical renormalization
technique to study the quench dynamics of Model J (given by Eq.\
(\ref{eq:dye}) at the critical point). Though this section does not contain
anything new of a fundamental nature, it does however compute exponents to
all orders in perturbation.

We first demonstrate that the precession term is relevant for the
quench dynamics to $T_c$. We will then calculate the $z$ and $\lambda$
exponents at this new fixed point. We will show, {\it to all orders in
perturbation}, that $\lambda$ is exactly equal to the spatial
dimension $d$. This latter fact, a consequence of the conservation law
(and indeed true for Model B dynamics too), may also be arrived at by
the general arguments presented in Ref.\ \cite{SM}.

In the absence of the torque term, the nontrivial fixed point is given by
the Wilson-Fisher value (WF), $u^{*}=(8/11)\pi^2\epsilon$, where
$\epsilon = 4 - d$. Power counting shows that the scaling dimension of
$g$ is $d/2 + 1 - z + \eta/2$, where the exponents take their WF values
$z=4-\eta$ and $\eta = (5/242) \epsilon^2$. This implies that the torque
$g$ is relevant at the WF fixed point when $d<6$ \cite{MA}.

We now have to determine this new torque driven fixed point and
calculate the dynamical exponents $z$ and $\lambda$. Both these
exponents can be obtained readily using general arguments, which we
briefly discuss. At the new fixed point it is clear that $g$ does not
get renormalized, which implies that $z=(d+2+\eta)/2$. Thus a
calculation of $z$ within perturbation theory reduces to a
calculation of $\eta$ at this fixed point \cite{MA}. Likewise
$\lambda$ can be obtained from the general arguments outlined in
Ref.\ \cite{SM}. A crucial ingredient in this argument (valid only for
quenches to $T_c$) is the demonstration that $S(k,t)$ obeys a scaling
form at $k=0$, a feature that was proved in Ref.\ \cite{KISS} to all
orders in perturbation for Model B dynamics. Here we {\it directly}
calculate both $z$ and $\lambda$ using diagrammatic perturbation
theory, and show that $\lambda =d$ to all orders in perturbation.

This is done 
within the Martin-Siggia-Rose (MSR) formalism \cite{JANS}. For our problem,
the MSR generating functional is,
\begin{eqnarray}
\lefteqn{{\cal Z}[\vec{h},\vec{\tilde{h}}] = \int{\cal
D}(\vec{\tilde{\phi}}){\cal
D}(\vec{\phi})\,\,
\exp\,\bigg\{\,-J[\vec{\phi},\vec{\tilde{\phi}}]-H_{0}[\vec{\phi}_{0}]
} \nonumber \\
                                &   & +\int_{0}^{\infty}dt \int 
d{\bf k}\,(\vec{\tilde{{h}}}_{\bf k} \cdot\vec{\tilde{\phi}}_{-\bf k}+
\vec{h}_{\bf k}\cdot\vec{\phi}_{-\bf k})\, \bigg\}
\label{eq:genfun}
\end{eqnarray}
with the MSR action written as
\begin{eqnarray}
\lefteqn{J[\vec{\phi},\vec{\tilde{\phi}}] = \int_{0}^{\infty}dt\int d{\bf k}
\bigg\{\vec{\tilde{\phi}}_{\bf
k}\cdot\bigg[\partial_{t}\vec{\phi}_{\bf k}+\Gamma k^2\frac{\delta
F[\vec{\phi}]}{\delta \vec{\phi}_{-\bf k}}
} \nonumber \\
                                      &  & +\int d{\bf
k}_1\bigg(\frac{g\Gamma}{2}(k_{1}^{2}-({\bf k}-{\bf k}_{1})^2)
\vec{\phi}_{{\bf k}_1}\times\vec{\phi}_{{\bf k}-{\bf k}_1}\bigg)
\bigg] \nonumber \\ 
                                     &  &
 -\Gamma k^2\vec{\tilde{\phi}}_{\bf  k}\cdot\vec{\tilde{\phi}}_{-\bf
k}\bigg\} \,\,.
\end{eqnarray}
In the expression for the generating functional, the initial distribution 
of the order parameter (gaussian with the width $ = \tau^{-1}_{0}$)
enters the form of  $H_{0} = \int d{\bf
k}\frac{\tau_{0}}{2}(\vec{\phi}_{\bf k}(0)\cdot\vec{\phi}_{-\bf
k}(0))$ \cite{JANS}.

Power counting reveals the presence of
two different upper critical dimensions coming from the 
quatric term ($d_c^{u}=4$) and the cubic torque term ($d_c^{g}=6$) in 
the action $J$. This implies we have to
evaluate the fixed points and exponents in a double power series expansion
in $\epsilon=4-d$ and $\varepsilon=6-d$ \cite{MA}. 

The unperturbed correlation $C^{0}_{\bf k}(t_1,t_2)$ and response
$G^{0}_{\bf k}(t_1,t_2)$ functions, and the bare $u$ and $g$ vertices are
shown in Fig.\ 20.  Again power counting shows that at $d=3$, our
perturbation expansion does not generate additional terms other than those
already contained in $J$, i.e. the theory is renormalizable. However the
perturbation theory gives rise to ultraviolet divergences which can be
removed by adding counter-terms to the action.

To remove these divergences,
we introduce renormalization factors
(superscripts $R$ and $B$ denote renormalized 
and bare quantities respectively),
$\vec{\tilde\phi}^{R}_{\bf k}(0)= (\tilde ZZ_0)^{-1/2}
\vec{\tilde\phi}^{B}_{\bf k}(0)$, $ \vec\phi^{R}_{\bf
k}(t) = Z^{-1/2} \vec\phi^{B}_{\bf k}(t)$,
$\vec{\tilde \phi}^{R}_{\bf k}(t)  =   {\tilde Z}^{-1/2}
\vec{\tilde \phi}^{B}_{\bf k}(t) $,
$u^{R} =  Z^{-1}_{u}u^{B}$,
$g^{R} = Z^{-1}_{g}g^{B}$,
$\Gamma^{R} = Z^{-1}_{\Gamma}\Gamma^{B}$ and 
$\tau^{R}_{0} = Z^{-1}_{\tau_0} \tau^{B}_{0} $.

Since the dynamics obeys detailed balance, the
renormalization factors $Z$ and $Z_{u}$ are the same as in
statics. Further the conservation of the order parameter forces $Z\tilde Z=1$ to all
orders.

The new fixed point is given by the zeroes of the $\beta$ functions of the
theory. The $\beta$ functions, calculated from the $Z$ factors, 
get contributions from all diagrams
containing the primitively divergent diagrams $\Gamma^{(2)}_{\phi 
\tilde\phi}$, $\Gamma^{(3)}_{\phi \phi \tilde\phi}$ and 
$\Gamma^{(4)}_{\phi \phi \phi \tilde\phi}$ (Fig. 20).

The new fixed point, to one loop, is given by
$g^{*}=\pm \sqrt{192\pi^3\varepsilon} + {\cal O}(\varepsilon^{3/2})$, 
$u^{*}=(8/11)\pi^2\epsilon + {\cal O}(\epsilon^2)$ (note
$u^{*}$ does not change from its WF value to all loops) and the dynamical 
exponent $z=4-\varepsilon/2+{\cal O}(\epsilon^2)$ \cite{MA}. 

\begin{figure}
\centerline{\epsfig{figure=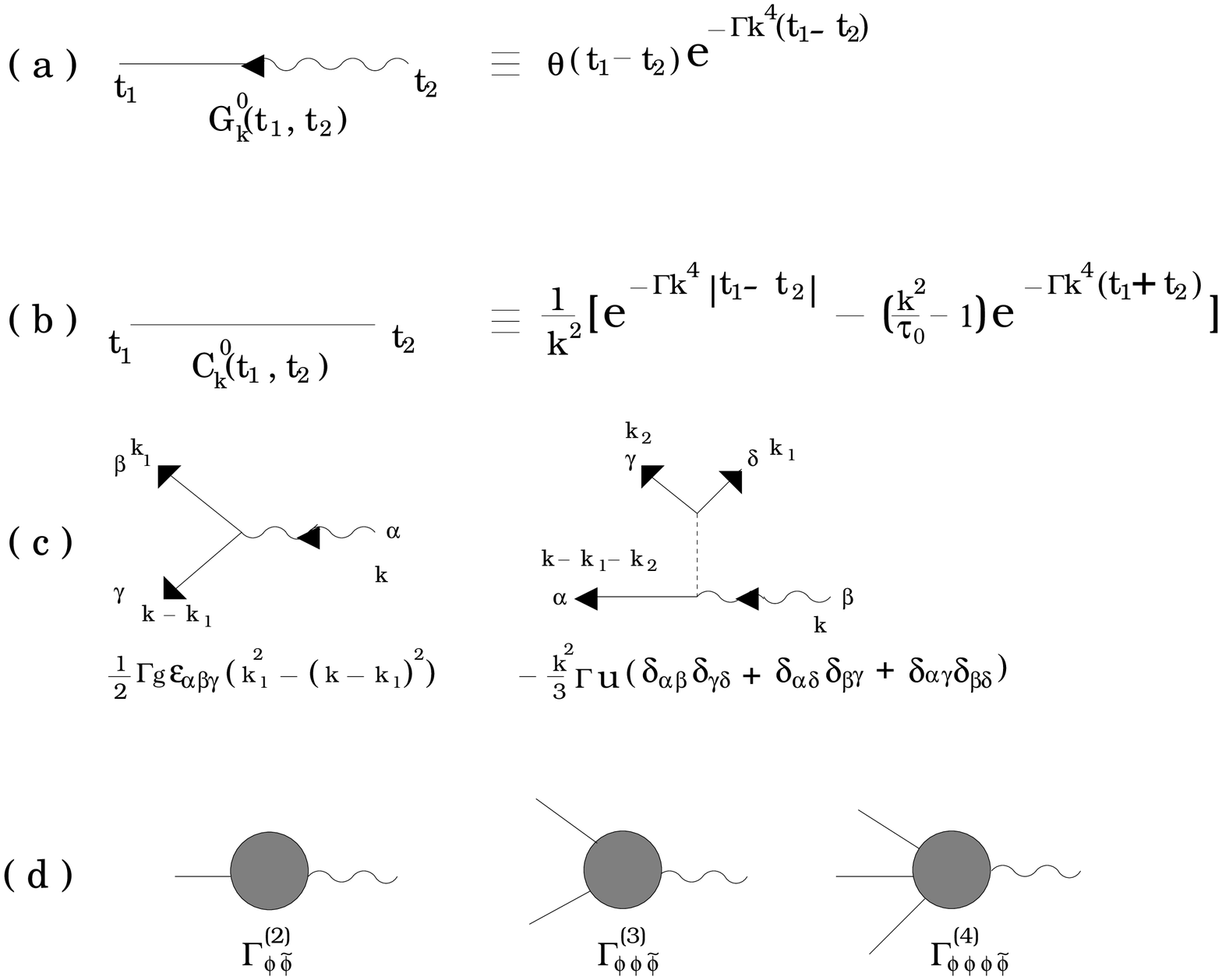,width=7.0cm,height=7.0cm}}
\end{figure} 

FIG.20. Unperturbed (a) response function $G^{0}_{\bf k}$, (b)
correlation function $C^{0}_{\bf k}$, and the (c) two bare vertices $u$ and $g$.
Wavy and straight lines represent the $\vec{\tilde{\phi}}_{\bf
k}(t)$ and $\vec{\phi}_{\bf k}(t)$ fields respectively. (d) Primitively
divergent diagrams $\Gamma^{(2)}_{\phi 
\tilde\phi}$, $\Gamma^{(3)}_{\phi \phi \tilde\phi}$ and 
$\Gamma^{(4)}_{\phi \phi \phi \tilde\phi}$.

\vskip.3cm

The $\lambda$ exponent can be computed from the response 
function $G_{\bf k}(t,0) \equiv \langle {\vec{\tilde
\phi}}_{\bf k}(0) \cdot {\vec \phi}_{-\bf k}(t) \rangle$ since this is 
equal to the autocorrelation function
${\tau_0}^{-1} \langle {\vec \phi}_{\bf k}(t) \cdot {\vec \phi}_{-\bf
k}(0) \rangle$, as can be seen from the first term in $J$ on integrating
by parts. The response function gets renormalized by

\begin{equation}
G^{R}_{k}(t,0)=Z_0^{-1/2} G^{B}_{k}(t,0)\,.
\end{equation}

The divergent contributions to $G_{B}$ could come from two sources. Each
term in the double perturbation series could contain the 
primitively divergent subdiagrams
$\Gamma^{(2)}$, $\Gamma^{(3)}$ or $\Gamma^{(4)}$, which we have already
accounted for by replacing these by 
their renormalized counterparts.
The other divergent contribution could arise
from the primitive divergences of the 1-particle reducible vertex function
$\Gamma^{(2)}({\bf k}, t, 0)$, defined by $G_{\bf k}(t,0) \equiv \int G_{\bf
k}(t-t')\,\Gamma^{(2)}({\bf k}, t', 0)\, d\,t'$. The superficial 
divergence of the diagrams contributing to $G_{\bf k}(t,0)$ is
$D=V_u(d-4)+\frac{V_g}{2}(d-6)-2 $ (where $V_u\,(V_g)$ is the number of $u$
($g$) vertices respectively). This is negative for all $d$. For
(a) when $d>6$, the only stable fixed point is the gaussian fixed
point and so $D=-2$, (b) when $4< d \leq 6$, $u$ is irrelevant and so
$D=\frac{V_g}{2}(d-6)-2 < 0$ and (c) when $d \leq 4$, $D$ is clearly
negative. This implies 
that $G^{B}_{\bf k}(t,0)$ does not get renormalized and $Z_0=1$.
Consequently
$\lambda$ stays at its mean-field value of $d$ for
this conserved Heisenberg dynamics both with and without the torque.

\section{Conclusions}

Traditional analysis of the asymptotic ordering dynamics of vector order
parameters focuses on the dynamics of defects, and ignores the bulk
excitations like spin waves which most often decay faster. In this work we
have looked at the very realistic model of Heisenberg spins with
precessional dynamics and have shown that the longer lived spin waves
couple to the defects even at late times, driving the system to a new
fixed point. The new `torque-driven' fixed point characterized by $z=2$
and $\lambda \approx 5.05$, is accessed after a crossover time $t_c \sim
1/g^2$ (where $g$ is the strength of the torque). Crossover scaling forms
describe physical quantities like domain size and equal/unequal time
correlation functions for all values of $g$. In the absence of the torque,
the spin-waves decay faster and so do not contribute to the asymptotic
dynamics.

We also studied the effects of the torque on the dynamics following a
quench to the critical point $T_c$. The torque is relevant with exponents
$z=4-\varepsilon/2$ and $\lambda = d$ (where $\varepsilon = 6-d$). We
found to all orders in perturbation theory that $\lambda =d$ which
follows as a consequence of the conservation of total magnetization.

We hope we have provided strong evidence that in order to go beyond the
present approximate theories of the asymptotic dynamics of conserved order
parameters, we need to systematically evaluate contributions coming from
the interaction of defects with spin-waves.

We thank D. Dhar for interesting discussions and Y. Hatwalne for a
critical reading of the manuscript.

\end{document}